\documentclass{aa}
\newcommand{\ha}{hereafter }
\newcommand{\xr}{X-ray }
\newcommand{\xrs}{X-rays }

\newcommand{\co}{Class~I }
\newcommand{\cd}{Class~II }
\newcommand{\ct}{Class~III }
\newcommand{\ro}{$\rho$ }
\newcommand{\TT}{T~Tauri }
\newcommand{\Einstein}{{\it Einstein Observatory }}
\newcommand{\ASCA}{{\it ASCA }}
\newcommand{\ROSAT}{{\it ROSAT }}
\newcommand{\ISO}{{\it ISO }}
\newcommand{\ISOCAM}{{\it ISOCAM }}
\newcommand{\PSPC}{{\it PSPC }}
\newcommand{\HRI}{{\it HRI }}

\def\ltsima{$\; \buildrel < \over \sim \;$} 
\def\gtsima{$\; \buildrel > \over \sim \;$}
\def\simlt{\lower.5ex\hbox{\ltsima}}
\def\simgt{\lower.5ex\hbox{\gtsima}}

\def\revtex@ver{2}    		
\def\revtex@date{21 March 00} 	
\def\revtex@jnl{A\&A}           
\def\revtex@genre{manuscript}   

\usepackage{graphics}

\begin{document}

\thesaurus{05
          (	10.152 \ro Ophiuchi cluster;
		08.16.5; 
    		08.6.2;
		13.25.5;
		13.09.6)}

\title{X-rays and regions of star formation:\\
a combined ROSAT-HRI/near-to-mid IR study \\
of the $\rho$~Oph dark cloud\thanks{Table A1, Fig. A1, Table B1 and Fig. B1 are only available 
in the on-line edition of the Journal (Table B1 is also available at the CDS).}}

\author{N. Grosso\inst{1, 2}, T. Montmerle\inst{1}, S. Bontemps\inst{3}, P. Andr{\'e}\inst{1}, and E.D. Feigelson\inst{4}}

\institute{Service d'Astrophysique, CEA/DSM/DAPNIA,  CEA-Saclay,
 F-91191 Gif-sur-Yvette Cedex, France \and
Max-Planck-Institut f{\"u}r extraterrestrische Physik, Giessenbachstra{\ss}e 1, D-85740 Garching, Germany \and
Observatoire de Bordeaux, BP 89, F-33270 Floirac, France \and 
Departement of Astronomy and Astrophysics, 
Pennsylvania State University,
University Park, PA 16802, USA}

\offprints{N.\,Grosso,\,ngrosso@xray.mpe.mpg.de}
\date{Received  / Accepted }

\titlerunning{A combined ROSAT-HRI/near-to-mid IR study of the $\rho$ Oph dark cloud}
\authorrunning{Grosso et~al.}

\maketitle
\begin{abstract}
We have obtained two deep exposures of the \ro Oph cloud core region 
with the \ROSAT {\it High Resolution Imager}.  
The improved position accuracy (1$\arcsec$--6$\arcsec$) with respect to previous
recent \xr observations (\ROSAT {\it PSPC}, and {\it ASCA}\,) allows us to remove
positional ambiguities for the detected sources. We also cross-correlate
the X-ray positions with IR sources found in the {\it ISOCAM} survey 
of the same region at 6.7 and 14.3\,$\mu$m, in addition to sources (optical and IR)
known from ground-based observations, which are young stars (T~Tauri
stars, with and without circumstellar disks, and protostars).
We thus obtain the best-studied sample of X-ray emitting stars in a
star-forming region (63 \xr sources detected, and 55 identified).

We find that there is no statistically significant difference between the X-ray
luminosity functions of {\it HRI}-detected Class~II and Class~III sources,
i.e., T~Tauri stars with and without disks, confirming
that the contribution of these disks to X-ray emission (for instance by
magnetic reconnection between the star and the disk), or to X-ray absorption, must be small.

X-ray variability of T~Tauri stars can be studied by comparing the \HRI data
with the previously obtained \PSPC data, but also using the fact that some \HRI
observations were done at different epochs.  The resulting statistics show that
most of the sources are variable, and that their variability is consistent with a
solar-like (hence magnetic) flare origin.

We use the information given both by the \ISOCAM survey and by our \HRI deep
exposure to study the T~Tauri star population of the \ro Oph dense cores.  
We confirm that essentially all Class~II and Class~III sources
(embedded T~Tauri stars) are X-ray emitters, and that a strong 			
correlation exists between their X-ray luminosity, $L_{\mathrm X}$, and their
stellar luminosity, $L_\star$, with $L_{\mathrm X}/L_\star \sim 10^{-4}$.
Most of the new \ISOCAM Class~II sources are not detected, however, which we explain
by the fact that their X-ray luminosities ``predicted'' on the basis of this
correlation are too faint to be detected by the {\it HRI}.  

We predict that $\sim$40 unknown faint or embedded Class~III sources
remain to be discovered in X-rays in the {\it HRI/ISOCAM} overlapping area,  
down to a limit of L$_{\mathrm{X}} \sim 3 \times
10^{28}$\,erg\,s$^{-1}$.  We show that the bulk of these unknown Class~III
sources should be made of low- to very low-mass stars (M$_\star <0.1$--0.6\,M$_\odot$).  
Prospects for future detections with {\it XMM-Newton} and {\it Chandra} are discussed.

\keywords{(Galaxy:) open clusters and associations: individual: $\rho$~Ophiuchi cluster
 -- stars: pre-main sequence -- stars: formation -- X-rays: stars -- infrared: stars}

\end{abstract}

\section{Introduction}

The \ro Ophiuchi dark cloud complex is one of the nearest active site of
low-mass star formation (see Wilking \cite{wilking92} for a review).
It is composed of two main dark clouds, L1688 and
L1689, from which filamentary dark clouds, called streamers, extend to the north-east 
over tens of parsecs (e.g., Loren \cite{loren89}; de Geus et~al. \cite{degeus90}).  The main star
formation activity is observed in the westernmost dark cloud, L1688, which shows a
rich cluster of low mass young stellar objects (YSO) around two dense
molecular cores, ``core~A'' and ``core~F'' in the terminology of Loren (\cite{loren89})
and Loren et~al. (\cite{loren90}). 

The distance to the molecular complex remains somewhat controversial (see Wilking \cite{wilking92}), 
with a usually adopted distance $d \sim 160$ pc from the Sun. 
From {\it Hipparcos} parallaxes and Tycho B--V colors of classes V and III stars,
Knude \& H$\o$g (\cite{knude98}) have detected at $d = 120$\,pc an abrupt rise of the reddening as
expected from a molecular cloud. 
Based on the {\it Hipparcos} positions, proper motions, and parallaxes, de Zeeuw et~al. (\cite{dezeeuw99}) gives
$d = 145 \pm 2$\,pc for the mean distance of the Upper Scorpius OB association.
We adopt $d =140$\,pc in this article, instead of 160\,pc used in our previous work.  

From infrared (IR) observations of star-forming regions, Lada and
collaborators (e.g.  Lada \cite{lada87}; Wilking, Lada, \& Wilking \cite{wilking89}, hereafter WLY)
introduced an IR classification and distinguished different stages of evolution
of young stellar objects (YSO).  This classification was subsequently
revisited by Andr{\'e} \& Montmerle (\cite{AM}, \ha AM) to incorporate results of
millimeter continuum studies on circumstellar dust.  The IR sources are
classified in three classes, according to their spectral energy distributions
(SEDs).  This classification, initially defined empirically, is now well
understood in terms of evolution of low-mass stars at their earliest stages.
Submillimeter observations led to the discovery of cold objects, younger
than the IR sources, and thus to the introduction of a fourth class named
``Class 0'' (Andr{\'e} et~al.  \cite{andre93}, \cite{andre00}).  Class 0 sources are very young
protostars, peaking in the submillimeter range, at the beginning of the main
accretion phase.  \co sources are evolved IR protostars, optically invisible, in
the late accretion phase.  \cd sources are YSO surrounded by optically thick
circumstellar disks.  \ct sources are YSO with an optically thin circumstellar
disk or no circumstellar disk.
Studies of optically visible YSO, T~Tauri stars,
led to another classification based on the H$_{\alpha}$ line, which separates
``classical'' T~Tauri stars (CTTS) from ``weak-line'' T~Tauri stars (WTTS)
according to their equivalent width in emission, with a boundary at
EW[H$_{\alpha}$]$\sim 5-20$\,\AA, depending on the spectral type (Mart{\'\i}n
1997).  CTTS and WTTS are usually taken to be identical to Class~II and
Class~III sources respectively, on the basis of their IR SED (see AM for a
discussion about these two classifications).  We will associate in this
article Class~II (Class~III) sources with CTTS (WTTS).

Several ground-based near-IR surveys (e.g.Wilking et~al. \cite{wilking89}; Greene et~al. \cite{GWAYL}, \ha GWAYL; 
Barsony et~al. \cite{barsony97}, \ha BKLT; and references therein) discovered 
in a $\approx$1 square degree area around the densest regions 
(with survey completeness limit down to $K \sim 14$),
$\sim$100 low-luminosity embedded sources. 
More recently, the \ISOCAM camera on-board the {\it Infrared Space Observatory} 
satellite imaged a half square degree centered on L1688 in the mid-IR (LW2 and LW3 filters, 
respectively centered at 6.7\,$\mu$m and 14.3\,$\mu$m -- \ISOCAM central programme surveys by Nordh et~al.; 
see Abergel et~al. \cite{abergel96}), and recognized 68 new faint young stars
with infrared excess (Bontemps et~al. \cite{bontemps00}).

Near-IR spectroscopy has been used to determine spectral types of an increasingly large 
number of \ro Oph YSO (see the pioneering works of Greene \& Meyer \cite{greene95}, and Greene \& Lada \cite{greene96}).
Recently, Luhman \& Rieke (\cite{luhman99}) obtained $K$-band spectroscopy for $\sim$100 sources,
combining a magnitude-limited sample in the cloud core ($K \le 12$) with a representative 
population from the outer region of the cluster ($K \le 11$).

The \ro Oph dark cloud YSO have also been extensively studied in X-rays.  Early observations
with the \Einstein satellite showed that at the \TT star stage YSO are bright
and variable \xr emitters in the 0.2--4\,keV energy band (Montmerle et~al. \cite{montmerle83}).
When the S/N ratio is sufficient large, their \xr spectra can be fitted by a
thin thermal model, with temperatures $\approx 1$\,keV and absorption column
densities $N_H \sim 10^{20}$--$10^{22}$\,cm$^{-2}$.  Variability studies and
modeling led to explain the \xr emission in terms of bremsstrahlung from a hot
($T_{\mathrm X} \sim 10^7$\,K) plasma trapped in very large magnetic loops, in other words in
terms of an enhanced solar-like flare activity (see reviews by Montmerle et~al.
\cite{montmerle93}; and Feigelson \& Montmerle \cite{FM}, \ha FM).  Casanova et~al. (\cite{CMFA}) -- \ha CMFA --
reported deep \ROSAT {\it Position Sensitive Proportional Counter} ({\it PSPC}\,) imaging of
the \ro Oph cloud dense cores A and F. They detected in the $35\arcmin \times 35\arcmin$ 
central portion of the field (the inner ring of  the \ROSAT detector entrance 
window support structure) 55 X-ray sources in the 1.0--2.4\,keV energy band.  
For three \xr sources, one or several Class~I sources lie within the error boxes 
of \xr peaks, 
but other counterparts are possible (unclassified IR sources, \TT stars).  \xr emission
from one of these Class~I sources, \object{YLW15} (=\object{IRS43} in WLY),  was unambiguously confirmed with a follow-up 
\ROSAT {\it High Resolution Imager} ({\it HRI}\,) observation by Grosso et~al. (\cite{grosso97}).  The
outer portion of the CMFA \PSPC field, analyzed by Casanova (\cite{casanova94}), contains 36 \xr
sources.  The optical spectroscopic classification of these \xr sources and
other \xr selected stars in the \ro Oph dark cloud vicinity, based on H$_{\alpha}$ 
and \ion{Li}{I} (670.8\,nm) spectroscopy, was made by Mart{\'\i}n
et~al. (\cite{martin98}), doubling the number of PMS stars
spectroscopically classified in the \ro Ophiuchi area.

Observations of harder \xr  ($>$4\,keV) from the \ro Oph dark cloud were initially only possible 
with non-imaging instruments.  {\it Tenma} and {\it Ginga}
revealed unresolved emission from the cloud core region, with a hard \xr spectrum
with $kT_{\mathrm X}\sim4$\,keV and $N_H\sim10^{22}$\,cm$^{-2}$ (Koyama \cite{koyama87}; Koyama et~al.
\cite{koyama92}).  Wide-energy band imaging observations became possible with \ASCA in the
range 0.5--10\,keV.  In the \ro Oph dark cloud, Koyama et~al. (\cite{koyama94}) detected hard 
\xrs from T~Tauri stars, with $kT_{\mathrm X}$ up to $\sim$8\,keV in the case of the WTTS DoAr21.  
There is also some evidence for unresolved hard \xr emission
from embedded young stars below the point source detection limit.
From this \ASCA observation,
Kamata et~al. (\cite{kamata97}), found additional \TT stars and detected three \xr sources associated with Class~I 
sources, but with large \xr error boxes (15$\arcsec$--30$\arcsec$).

There is a deep connection between IR and \xr observations of star-forming regions.
Sensitive ground-based near-IR surveys penetrate dark clouds (except for
dense cores) so that their 
source populations are frequently dominated by ordinary stars in the Galactic disk.  
Space-based mid-IR isolates YSO with significant 
circumstellar material and effectively eliminates the background 
star population, but they will miss the recognition of YSO with less massive or 
absent disks.  X-ray emission, in contrast, is elevated by 1--4 orders of 
magnitude in YSO of all ages, irrespective of a disk presence.  
It thus provides a unique tool for improving the census of young star clusters. 

In this article, we present the results from the \HRI follow-up
of the CMFA \PSPC observation.  
The high angular resolution of these observations allows us to
find counterparts to all \xr sources without ambiguity.  The comparison with
the sensitive \ISOCAM survey of the \ro Oph dark cloud significantly improves the
existing classification of these counterparts and allows us to do statistical studies on
a well defined sample.

We first present the \ROSAT \HRI observations: image analysis, source detection
and identification ($\S$\ref{X-ray}).  
We incorporate the \ISOCAM survey results from Bontemps et~al. (\cite{bontemps00}) 
and we present the resulting IR classification for the \HRI sources ($\S$\ref{Nature}).  
The next sections discuss the \xr luminosity of the \HRI detected TTS 
($\S$\ref{Luminosity}), and the X-ray detectability of the embedded TTS population ($\S$\ref{properties}).
Next (\S\ref{origins}), we show that the \HRI census of Class~III sources cannot be
complete, and that numerous unknown low-luminosity Class~III sources, perhaps 
including brown dwarfs, must exist.
Summary of the main results and conclusions are presented in $\S$\ref{Summary},
where prospects for improvements with {\it XMM-Newton} and {\it Chandra}, are also discussed.

Appendix~A gives details about the \HRI \xr source detection, and lists the \xr detections.
Optical finding charts, and identification list of the \HRI \xr sources can be found in Appendix~B.
Appendix~C compares these \HRI observations with previous \PSPC ones.
Appendix~D discusses the status of optical/IR counterparts without IR classification.

\begin{table*}[!htb]
\small
\caption{Log of \ROSAT \HRI Observations.}
\label{tab:log}
\begin{tabular}{lcclrlrrrr}
\hline
\hline
\vspace{-0.3cm}\\
&   &    & \multicolumn{2}{c}{Begin} & \multicolumn{2}{c}{End} & \multicolumn{1}{c}{Elapsed} \\
\vspace{-0.4cm}\\
\multicolumn{1}{c}{Pointing} & \multicolumn{1}{c}{Core} & \multicolumn{1}{c}{Obs.}  &\multicolumn{2}{c}{\hrulefill}&\multicolumn{2}{c}{\hrulefill}& \multicolumn{1}{c}{Time} & \multicolumn{1}{c}{Exposure}\\
\multicolumn{1}{c}{ID}   &  & \multicolumn{1}{c}{\#}  & \multicolumn{1}{c}{Date} & \multicolumn{1}{c}{Hour (UT)} & \multicolumn{1}{c}{Date} & \multicolumn{1}{c}{Hour (UT)} & \multicolumn{1}{c}{[ks]} & \multicolumn{1}{c}{[ks]} \\
\hline
201835h   & A & 1  & 1995 Aug 29 & 18$^{\mathrm h}$31$^{\mathrm m}$30$^{\mathrm s}$ & 1995 Sep ~12 & 22$^{\mathrm h}$35$^{\mathrm m}$47$^{\mathrm s}$ & 1\,224.3 & 51.3\\
\hline
201834h   & F & 1  	& 1995 Mar 09  & 00$^{\mathrm h}$51$^{\mathrm m}$03$^{\mathrm s}$ & 1995 Mar 14 & 05$^{\mathrm h}$16$^{\mathrm m}$43$^{\mathrm s}$ & 447.9   & 12.5 \\
201834h-1 & F & 2  	& 1995 Aug 18  & 19$^{\mathrm h}$29$^{\mathrm m}$21$^{\mathrm s}$ & 1995 Aug 20 & 14$^{\mathrm h}$52$^{\mathrm m}$05$^{\mathrm s}$ &  27.8   & 27.5 \\
201834h-2 & F & 3  	& 1996 Sep ~07 & 23$^{\mathrm h}$12$^{\mathrm m}$46$^{\mathrm s}$ & 1996 Sep ~11& 13$^{\mathrm h}$46$^{\mathrm m}$23$^{\mathrm s}$ & 311.6   & 37.2 \\
          & F & 1+2+3  & 1995 Mar 09  & 00$^{\mathrm h}$51$^{\mathrm m}$03$^{\mathrm s}$ & 1996 Sep ~11& 13$^{\mathrm h}$46$^{\mathrm m}$23$^{\mathrm s}$ & 47\,739.3 & 77.2 \\
\hline
\hline
\end{tabular}
\end{table*}

\section{The ROSAT HRI X-ray observations}
\label{X-ray}

	\subsection{The ROSAT \HRI images}

\begin{figure}[!hbt] 
\resizebox{\hsize}{!}{\includegraphics{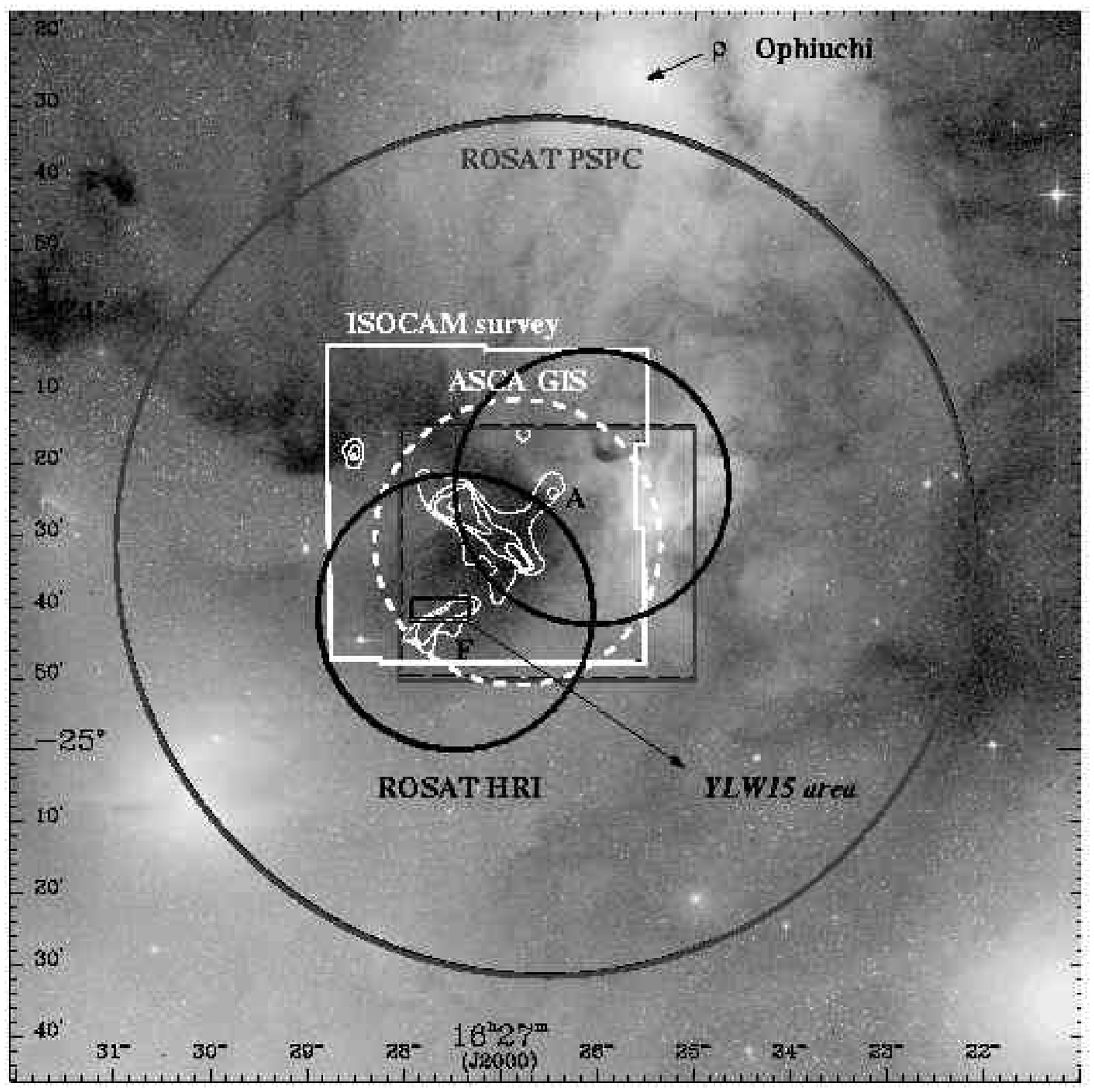}}
\caption{Map of the $\rho$ Ophiuchi dark cloud observation fields.
2\degr diameter \ROSAT \PSPC field of view, and its central
$35\arcmin \times 35\arcmin$ studied by Casanova et~al. (\cite{CMFA});
40\arcmin diameter \ROSAT \HRI fields of view and \object{YLW15} area published in Grosso
et~al. (\cite{grosso97}); \ASCA {\it GIS} field of view from Koyama et~al. (\cite{koyama94}) and Kamata et~al. (\cite{kamata97}); \ISOCAM survey
field (central programme observations by Nordh et~al.; see Abergel et~al. \cite{abergel96}).  
The background image of the $\rho$ Ophiuchi star forming region is taken from the first Digitized Sky
Survey, contrast was enhanced with Gaussian histogram specification (Canzian \cite{canzian97}).  
Regions of high visual extinction are clearly visible.  DCO$^+$(J=2-1) contours from 
Loren et~al. (\cite{loren90}) show for reference the location of dense cores A, F.}
\label{fields-map} 
\end{figure}

We have observed the dense cores A and F of the \ro Oph dark cloud with the
\ROSAT {\it HRI}. The detector is sensitive to the 0.1--2.4\,keV energy range, but 
has no spectral resolution. 
The two observation fields were respectively centered on 
approximatively the WTTS \object{DoAr21} ($\alpha = 16^{\mathrm h}26^{\mathrm m}2\fs4,~\delta = -24\degr23\arcmin24\arcsec$
[J2000]) and on the Class~I protostar \object{YLW15} = \object{IRS43} ($\alpha = 16^{\mathrm h}27^{\mathrm m}26\fs4,~\delta =
-24\degr40\arcmin48\arcsec$ [J2000]).
\ROSAT \HRI images have a diameter of
$\sim 40\arcmin$.  Fig.~\ref{fields-map} displays the two observation fields,
which include the dense DCO$^+$ cores A, B, C, E, and F (Loren et~al. \cite{loren90}), 
most of the area studied by CMFA and {\it ASCA}, as well as
the \ISOCAM survey.  The first observation field,
centered approximatively on $\rho$~Oph~A, will be referred to as the ``core~A
field''; the second observation field, centered on dense core~F, will be called
the ``core~F field''.  The core~A field was observed between 1995 August 29 and
1995 September 12 with a total exposure of 51.3\,ks.  
The core~F field was observed at three different epochs (\ha observations \#1, \#2, and \#3):  
between 1995 March 9 and 14 (12.5\,ks), between 1995 August 18 and 20 (27.5\,ks), 
and between 1996 September 7 and 11 (37.2\,ks).  
These three different epochs give a total exposure of 77.2\,ks (see Table~\ref{tab:log} 
for the log of \ROSAT \HRI observations details).

	\subsection{The HRI image analysis}

We have analyzed separately the four data sets.  
Standard source detection algorithms were used to find \xr sources, 
and to search optical counterparts for each set.
We then selected \xr sources 
with the best position accuracy (usually the brightest not
too far away from the field center) and having an unambiguous counterpart,
corrected all \xr positions from the existing offsets, and used this improved
astrometry to remove possible ambiguities in the identification of the X-ray
sources. Details can be found in Appendix~A.

We find 63 \HRI \xr sources.
Fig.~\ref{Sources} shows the positions of these sources,
superimposed on a combined IR and optical image of the \ro Oph cloud.
Coordinates, error boxes, likelihoods of existence, 
and count rates of these \HRI \xr sources can be found in Appendix~A (Table~\ref{tab:sources}).

	\subsection{\HRI \xr source identifications}

Identification of the \HRI \xr sources was made by cross-correlations with published
lists of confirmed or suspected cloud members (AM), IR surveys (Greene \& Young 1992; BKLT; and 
Bontemps et~al. \cite{bontemps00}, see $\S$\ref{Nature}), $K$-band spectroscopic (Luhman \& Rieke \cite{luhman99}), 
radio surveys (Andr{\'e}, Montmerle, \& Feigelson \cite{andre87}; Stine et~al. \cite{stine88}; Leous et~al. \cite{leous91}), 
and with previously known \xr sources 
({\it PSPC}: CMFA, Mart{\'\i}n et~al. \cite{martin98}; {\it ASCA}:  Kamata et~al. \cite{kamata97}).  
For \xr sources without published counterparts we have used {\it SIMBAD},\footnote{On-line
version at {\it http://simbad.u$-$strasbg.fr/Simbad}\,.}  
and we have also searched optical counterparts on optical red band images from the Digitized Sky
Survey.\footnote{On-line version on the ESO site:  
{\it http://arch$-$http.hq.eso.\-org/cgi$-$bin/dss}\,.}

As shown by the finding charts in Appendix~B, thanks to the good angular resolution of the 
\ROSAT {\it HRI} ( PSF FWHM $\sim 5\arcsec$ on axis; due to the mirrors as well as the detector), 
the position accuracy ($1\arcsec$--$6\arcsec$, see Col.~6 Table~\ref{tab:sources}) 
allows us to find counterparts almost without ambiguity.\footnote{When two 
possible counterparts are in the error box (this happens only three times, 
see the finding charts in Appendix~B, and Table~\ref{tab:ident} notes), 
we take the more luminous in the $J$-band, 
since the \xr luminosity of TTS is correlated with the $J$-band luminosity 
(see CMFA, and below $\S$\ref{Cloud_extinctions}, $\S$\ref{Lstar-Lx}).}
Identification lists for the core~A and core F fields are given in Appendix~B, Table~\ref{tab:ident}.  
We also discuss in Appendix~B the identification of X-ray sources with a low statistical significance.

Nearly $90\%$ of the \HRI \xr sources are identified.
We detect only $\sim 70\%$ of the \PSPC \xr sources (CMFA; Casanova \cite{casanova94}), 
but this can be explained by the difference in sensitivity between the two instruments, 
and the intrinsic variability of the \xr sources (see  Appendix~C).

\section{Nature and IR properties of the \HRI sources}
\label{Nature}
	
Following the results of CMFA, essentially all X-ray sources we found in the
$\rho$~Ophiuchi dark cloud should be young stars, a number of them being still embedded
in the cloud. Embedded YSO are mainly studied at IR and millimeter wavelengths,
but they may all be potential X-ray emitter regardless of their IR classification. 
Indeed, using the results of Wilking et~al.  (\cite{wilking89}), AM, and GWAYL, CMFA analyzed their results 
in the light of the IR observations of the stars they observed in X-rays.  
We can then (i) use the published IR surveys to provide a list
of recognized YSO members of the cluster to be compared to the observed
X-ray properties, and (ii) use X-ray emission to discriminate between true cluster 
members and the many potential background stars seen in IR images.

\begin{figure*}[!ht] 
\resizebox{\hsize}{!}{\includegraphics{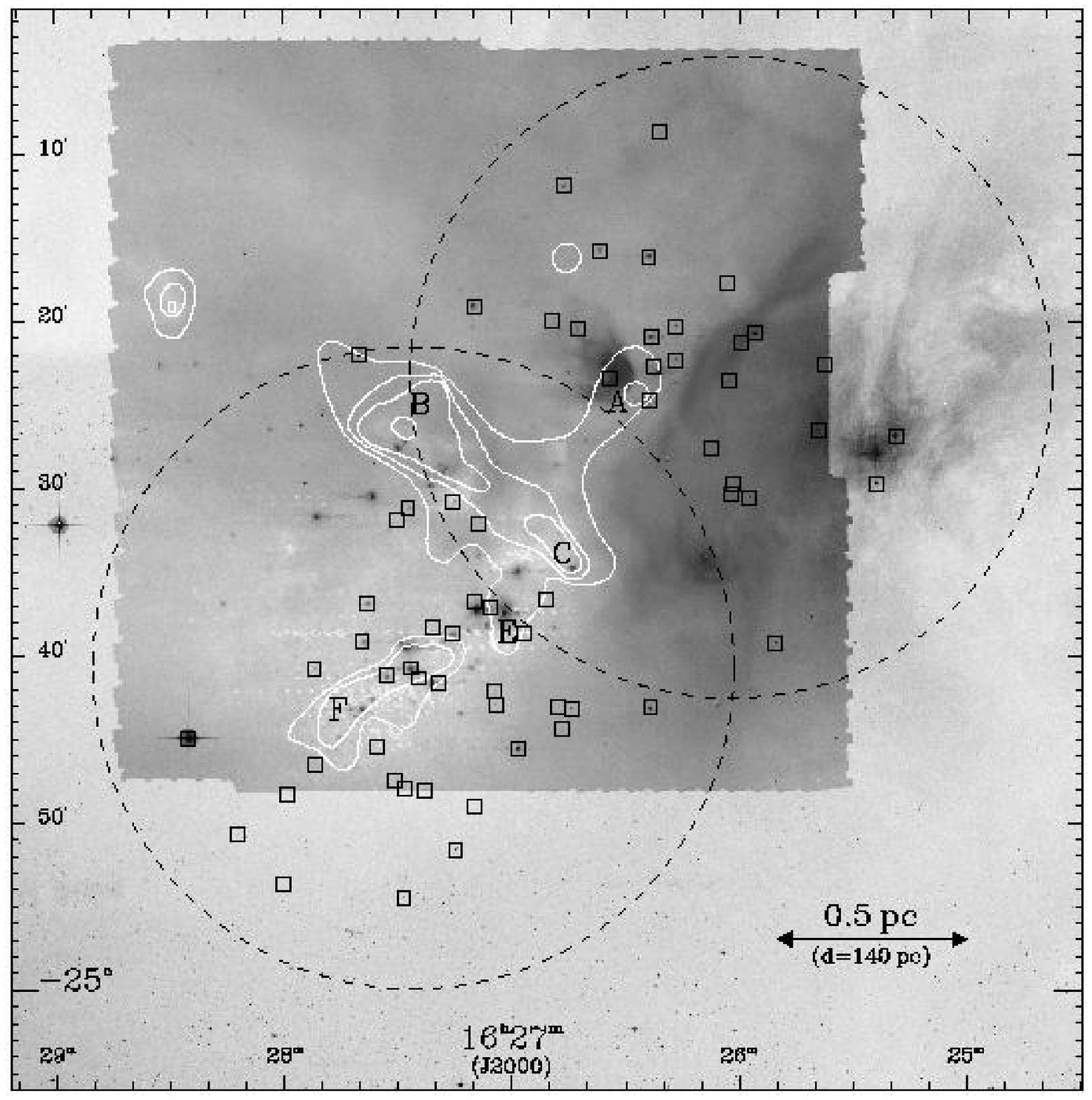}}
\caption{\xr sources in the \ROSAT \HRI fields.  The
composite \ISOCAM map of Abergel et~al. (\cite{abergel96})
(LW2 filter image [5--8.5\,$\mu$m], plus LW3 filter image [12--18\,$\mu$m]) is
merged with the background optical image taken from the first Digitized Sky Survey. 
DCO$^+$(J=2-1) contours show the location of dense cores named A, B, C, E, F by Loren et~al. (\cite{loren90}).
A scale of 0.5\,pc is shown for $d=140$\,pc.
The positions of the \xr sources are marked by $50\arcsec$-size squares.  
Typical X-ray error positions range between 1$\arcsec$--6$\arcsec$.}
\label{Sources}
\end{figure*}

	\subsection{New Class~II and Class~III source census after \ISOCAM}
	\label{New-ISOCAM-census}

Near-IR surveys of the \ro~Ophiuchi cluster, sensitive 
enough to detect low-luminosity embedded young stars, have recently been  
published (Comer{\'o}n et~al. \cite{comeron93}; Strom et~al. \cite{strom95} -- \ha SKS; BKLT).
However these ground-based surveys encountered limitations in
recognizing the nature of all the embedded sources, and have
therefore not much increased the number of {\it bona fide} members of the
\ro~Oph cluster.
The mid-IR camera aboard {\it ISO}, {\it ISOCAM}, 
produced a map of the $\rho$~Oph main cloud, 
used by Bontemps et~al. (\cite{bontemps00}) to study the young star population.  
This mid-IR study resulted in a
significantly more complete census of the $\rho$~Oph cluster population.  
We here use this new census as a basis for discussion about the nature of the detected X-ray
sources and to estimate the occurrence and properties of the X-ray emission
of the different classes of YSO.

The mid-IR photometry at 6.7 and 14.3\,$\mu$m appears invaluable to characterize
sources with IR excesses, i.e., Class~I sources and Class~II sources
(e.g., Nordh et~al. \cite{nordh96}; Bontemps et~al. \cite{bontemps98}).  In $\rho$~Oph, Bontemps
et~al. (\cite{bontemps00}) have doubled the number of Class~II sources known.  
These authors conclude that the sample is complete down to stellar
luminosities, $L_{\star}$, as low as 0.03\,L$_\odot$, thus
extending downwards the luminosity function obtained from the ground
by about one order of magnitude.  However, just as in the near-IR, these
measurements alone cannot characterize the nature of sources without IR excess:  
these can be either Class~III sources (diskless TTS) or background sources.

The tentative Class~III classification of a number of sources coming from
previous \xr observation (CMFA), or from this article, is now confirmed by {\it
ISOCAM}, which detected no IR excess.  We call these sources ``new'' Class~III
sources (see Table~\ref{tab:ident}, and $\S$\ref{TTS_sample}).

Since we already identified one of the \HRI sources as a foreground star
(\object{ROXF37} = \object{HD148352}, see Table~\ref{tab:ident}), the question arises that a
number of these new sources with Class~III spectra may also be field stars,
contaminating the genuine young star sample.  Guillout et~al.  (\cite{guillout96}) has
estimated the stellar content of flux-limited X-ray surveys, based on the
age-dependent stellar population model developed by the Besan{\c c}on group.
For the average sensitivity of our \HRI observation ($\sim 6 \times
10^{-4}$\,cts\,s$^{-1}$, see Fig.~\ref{upper}), the $\rho$~Oph galactic
latitude ($\sim 20\degr$), and our total field of view (0.5 square degrees), this
model yields an estimate of 5 contaminating stars (P.  Guillout, private
communication).  Removing \object{HD148352}, this leaves 4 possible field star
candidates among the 21 ``new'' Class~III sources and 
``Class~II-III'' sources listed in Table~\ref{tab:ident}.  This number is
small enough not to affect significantly our discussion, and in what follows
we will simply neglect the possible contamination of our various Class~III
samples by field stars.

The Class~III source population will be studied in detail below ($\S$\ref{Luminosity} and $\S$\ref{origins}).

	\subsection{Cloud extinctions and stellar luminosities from near-IR photometry}
	\label{Cloud_extinctions}

YSO suffer from large amounts of extinction by dust -- cloud dust plus circumstellar
dust -- (up to A$_{\mathrm{V}} = 60$ or more in \ro Oph), which strongly affects their fluxes 
at all wavelengths of interest, but does not necessarily prevent their detection in 
soft X-rays (see, e.g., CMFA).  
Despite this effect, it is possible to estimate the total extinctions 
along the line of sight and stellar luminosities.  
Near-IR photometry data appear to provide the most reliable estimate
for the luminosity of the embedded young stars (see discussion in Bontemps et~al.
2000 and references therein):  the $J$-band fluxes are usually almost purely
photospheric and thus trace the stellar luminosities very well (e.g., GWAYL; SKS).  
Similarly the $J-H$ color is a sensitive tracer of interstellar extinction.
We therefore use the extinctions and stellar luminosities derived from near-IR 
photometry for the ISOCAM sample by Bontemps et~al. (\cite{bontemps00}).

In CMFA the bolometric luminosities were approximately estimated from the $J$-band fluxes
using the GWAYL conversion based on an observational correlation between $J$ 
fluxes and bolometric luminosities for Taurus and Chameleon TTS. 
The bolometric luminosity comprises in principle the total luminosity
(accretion + stellar) of a YSO. 
However for young PMS stars like Class~II sources and Class~III sources, 
this luminosity should coincide with their stellar luminosity since their accretion
luminosity (if any) has become significantly smaller than the photospheric luminosity. 
Finally we note that the conversion between the $J$-band absolute magnitude 
and the stellar luminosity used by Bontemps et~al. (\cite{bontemps00}) is numerically similar 
to the CMFA conversion between the dereddened $J$-band and the bolometric luminosity.

Cols. 8--9 of Table~\ref{tab:ident} give for each source its more up-to-date values 
of interstellar extinction and stellar luminosity, determined by Bontemps et~al for 140\,pc.
The reader will find the details of the calculations in Bontemps et~al. (\cite{bontemps00}).

	\subsection{Nature of the \HRI sources from IR data}
	\label{nature}

The YSO evolutionary stage, inferred from IR spectral energy distributions 
(see $\S$\ref{New-ISOCAM-census}), is available for most of the X-ray sources.   
The resulting census of the 55 X-ray sources with 
stellar counterparts is: 
one Class~I protostar (\object{YLW15}=\object{IRS43}); 
23 Class~II sources, including 4 new \ISOCAM Class~II sources; 
21 Class~III sources, including 13 new Class~III sources\footnote{4 were already 
\PSPC Class~III source candidates, and are confirmed by {\it ISOCAM}, which detected no IR excess; 
3 were probably detected by the {\it PSPC}, but due to its lower angular resolution the optical or IR
counterpart was uncertain (see Col.~2 of Table~\ref{tab:ident} and attached
notes); 6 are genuine new \xr detections, probably resulting from variability.}; 
8 whose classification is either Class~II or Class~III sources (see discussion in 
Appendix~D); one Class~III early-type star 
(\object{S1}, with spectral type B3; see Andr{\'e} et~al. \cite{andre88}), and one main sequence 
foreground star (\object{HD148352} with spectral type F2V; see the {\it Hipparcos} 
catalogue).

In Table~\ref{tab:ident}, Col.~6 lists cross-identifications with the \ISOCAM survey: 
{\it red} ({\it blue}\,) means \ISOCAM sources with (without) IR excess. 
Col.~7 gives the IR classification, Cols.~8--9 the extinctions and
stellar luminosities (from Bontemps et~al. \cite{bontemps00}).

Therefore, the present analysis has revealed no new X-ray emitting Class~I source, 
apart from the Class~I protostar YLW15 which has been the subject of a specific study (Grosso et~al. \cite{grosso97}).
The {\it HRI}/\ISOCAM sources are overwhelmingly TTS, 
for which improved results are described in the following sections.

\section{X-ray luminosities of the {\it HRI}-detected T~Tauri stars}
\label{Luminosity}

	\subsection{Derivation from source count rates}
	\label{derivation_from_source_counts}

We assume a fiducial TTS X-ray spectrum (see Montmerle \cite{montmerle96}) having
$kT_{\mathrm X}$=1\,keV plasma, with cosmic abundances, and Raymond-Smith line emissivities, 
and with interstellar absorption based on Morrison \& McCammon (\cite{morrison83}) cross sections.  
We use the relation given by Ryter (\cite{ryter96}) to estimate the hydrogen column density, 
$N_H$, from the visual extinction, $A_{\mathrm{V}}$, determined from IR data\footnote{Noted 
$A_{\mathrm{V,IR}}$ in CMFA.}: $N_H=2.23 \times 10^{21}\,A_{\mathrm{V}}$\,cm$^{-2}$.
The intrinsic (i.e., extinction corrected) \xr luminosities in the full \ROSAT 
energy band (0.1--2.4\,keV), $L_{\mathrm X}$, were calculated for classified sources 
from source count rates for $d \sim 140$\,pc using EXSAS, and are given in Cols.~10--13 of 
Table~\ref{tab:ident}.
The intrinsic  \xr luminosities span the range 
$L_{\mathrm X} \sim 1 \times 10^{29}~$--$~\sim 4 \times 10^{31}$\,erg s$^{-1}$.
(For \xr sources without extinction estimates we label the detection exposure with a question mark.)

	\subsection{Luminosity functions of {\it HRI}-detected Class~II and Class~III sources}

We here compare the extinction corrected \xr luminosities of the Class~II and Class~III sources
detected by the \HRI in the \ISOCAM field\footnote{22 Class~II sources and 19 Class~III sources;
for the core~F observation we take the mean \xr luminosity in Col.~13 of Table~\ref{tab:ident}.} 
in order to evaluate the contribution of the circumstellar disk to the \xr absorption, 
or to the \xr emission (for instance by magnetic reconnection between the star and the disk; 
see Montmerle et~al. \cite{montmerle00}).

\begin{figure}[!ht]
\resizebox{\hsize}{!}{\includegraphics{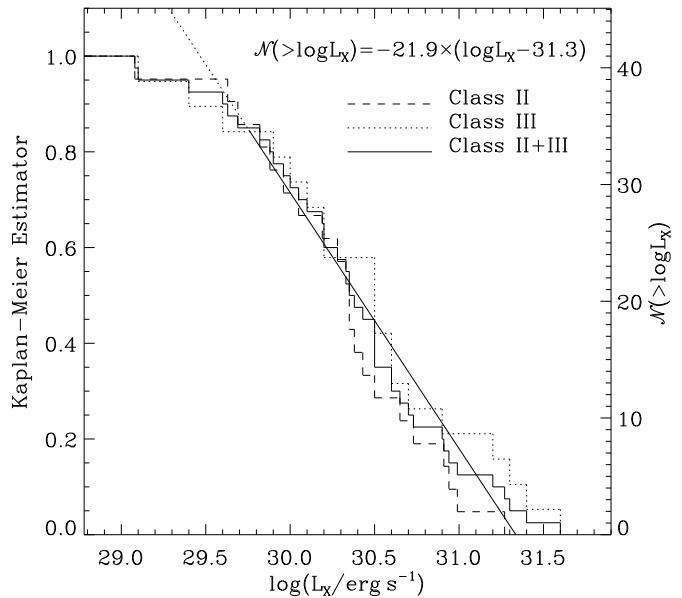}}
\caption{Cumulative normalized \xr luminosity functions of Class~II and 
Class~III sources detected with the \HRI in the \ro Oph cloud.
The dashed (dotted) histogram shows the Class~II (Class~III) source integral \xr luminosity 
functions.
The solid histogram shows the total cumulative \xr luminosity functions.
The straight line shows the fit for the total cumulative \xr luminosity functions.
The left scale gives the Kaplan-Meier estimator. 
The right scale gives the cumulative number of \xr sources.}
\label{luminosity-function}
\end{figure}

Fig.~\ref{luminosity-function} shows the cumulative \xr luminosity distribution
functions for these two populations, estimated using the ASURV statistical software
package (rev.~1.2;\footnote{Version available at 
{\it http://www.astro.psu.edu/statcodes}\,.} La Valley et~al. \cite{lavalley92}), 
which takes upper limits into account.  
These distributions are mathematically identical to the maximum likelihood 
Kaplan-Meier estimator (Feigelson \& Nelson \cite{feigelson85}).  
Their mean \xr luminosities (in erg\,s$^{-1}$) are given by 
$<\log(L_{\mathrm X})> = 30.3 \pm 0.1$ for Class~II sources,
and $<\log(L_{\mathrm X})> = 30.4 \pm 0.2$ for Class~III sources. 
We used nonparametric
two-sample tests implemented in ASURV --- Gehan's generalized Wilcoxon test,
Logrank test, Peto \& Peto generalized Wilcoxon test, Peto \& Prentice
generalized Wilcoxon test --- to see whether the difference between the two
luminosity functions is significant.  These tests gave a high probability
(46--72\,$\%$) that they are statistically indistinguishable.
This result agrees with previous deep studies of the \ro Oph main cloud (CMFA) and 
Chamaeleon I (Feigelson et~al. \cite{feigelson93}; Lawson et~al. \cite{lawson96}) YSO populations.

In contrast, Neuh{\"a}user et~al. (\cite{neuhaeuser95}) found in the Taurus-Auriga star-forming region
that Class~III sources are more \xr luminous than Class~II sources. 
Neuh{\"a}user et~al. (\cite{neuhaeuser95}) used the \ROSAT {\it All Sky Survey} ({\it RASS}\,) to cover
a large area of the Taurus-Auriga ($\sim 900$ square~degrees), 
including dense cores, but also away from this star-forming region. 
This shallow survey and large area must be compared to our deep pointed observations, 
where our field of interest covers 0.5 square~degrees.
In our observations we focus only on the dense cores, studying a younger population of YSO.
Contamination of the T~Tauri star sample by a more evolved, wide-spread, and older Class~III source
population, may explain the discrepancy with the result of Neuh{\"a}user et~al. (\cite{neuhaeuser95}).\footnote{In addition 
one should note that the {\it RASS} includes the soft band (0.1--1\,keV), which is more sensitive 
to the extinction than the hard band (1--2.4\,keV) taken by CMFA.}

Thus, we confirm that the contribution of the disk of Class~II sources to
their X-ray emission, or to X-ray absorption, must be small.

Next, we combine the \xr luminosity distributions of Class~II and Class~III sources to obtain the 
cumulative X-ray luminosity distribution function of all TTS detected by the {\it HRI}.  
For $\log(L_{\mathrm X})$ between 29.7 ($\log(L_{\mathrm{X,break}})$) and 31.3 ($\log(L_{\mathrm{X,max}})$), 
the distribution is loglinear:
${\cal N}(>\log(L_{\mathrm X}))=-21.9 \times (\log(L_{\mathrm X}) -31.3)$.
For $\log(L_{\mathrm X}) \le \log(L_{\mathrm{X,break}})$, the distribution shows a downwards trend, 
which is due to our lower efficiency to detect weak \xr emitting TTS.
We deduce from this linear relation the total \xr luminosity emitted by the X-ray 
sources with \xr luminosity between $L_{\mathrm{X,max}}$ and $L_{\mathrm{X,min}}$ (an arbitrary value):
$L_{\mathrm{X,tot}}= 21.9 \times (L_{\mathrm{X,max}}-L_{\mathrm{X,min}})$.
Thus, the total \xr luminosity of this group of X-ray sources is dominated by the 
brightest sources (with $L_{\mathrm{X,max}}$), and is very weakly dependent on the loglinear fit. 
For $L_{\mathrm{X,min}}=L_{\mathrm{X,break}}$, 
we find $L_{\mathrm{X,tot}} \sim 4.3 \times 10^{32}$\,erg\,s$^{-1}$.
This value is close to the asymptotic value $4.4 \times 10^{32}$\,erg\,s$^{-1}$ 
obtained taking $L_{\mathrm{X,min}}=0$, thus future detections of new \xr emitting TTS with low \xr luminosity 
will not greatly affect this result.

	\subsection{\HRI source variability}
	\label{Variability}

We present in Appendix~C a study of the variability 
of the \xr sources which were observed both by the \HRI and  the {\it PSPC}, 
and we show that some sources were in a high \xr state during the \HRI 
or the \PSPC observations. Here, we study
the variability of the  Core~F \HRI sources.

Core~F observations comprise three time-separated observations, which allow us
to reiterate the ``Christmas tree'' luminosity study made with \Einstein by
Montmerle et~al. (\cite{montmerle83}).  The idea suggested by the similarity between
Class~II sources and Class~III sources in X-rays was to assume that all the \xr sources
are basically one single type of \xr object, seen in different states.  The
result was that the distribution of the flux variations could be approximated by a
power-law with an index  $\beta = -1.4$.

\begin{figure}[!b] 
\resizebox{\hsize}{!}{\includegraphics{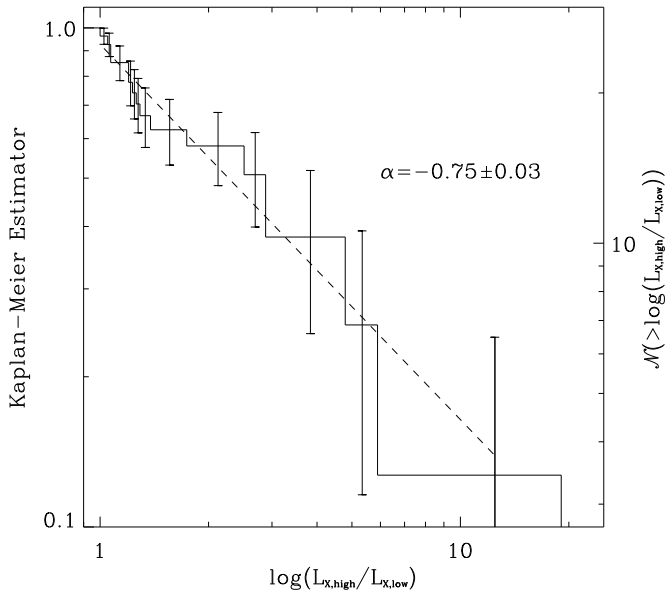}}
\caption{Distribution of X-ray luminosity variations.
We have estimated whenever possible for each \xr source of core~F the \xr variability
amplitude with the ratio high- on low-observed luminosity 
during the three observations.  
The frequency distribution has been estimated using the maximum likelihood
Kaplan-Meier estimator.  We find a power-law distribution:  
\( {\cal N}(>\frac{L_{\mathrm{X,high}}}{L_{\mathrm{X,low}}}) \propto (\frac{L_{\mathrm{X,high}}}{L_{\mathrm{X,low}}})^{\alpha}\), 
with $\alpha=-0.75\pm0.03$, consistent with a variability due to stellar flares, which 
supports the analogy with the solar magnetic activity.}
\label{power-law}
\end{figure}

We have estimated whenever possible for each Core~F \HRI source the \xr flux
variations from the observed high/low luminosity ratio,
${\cal R} \equiv L_{\mathrm{X,high}}/L_{\mathrm{X,low}}$, based on
 the three observations. This yields 27 values
including 11 lower limits.  
The integral distribution for a given ratio is
estimated using the maximum likelihood Kaplan-Meier estimator (see Fig.~\ref{power-law}).  
We find a power-law distribution:  
${\cal N}(>L_{\mathrm{X,high}}/L_{\mathrm{X,low}}) \propto (L_{\mathrm{X,high}}/L_{\mathrm{X,low}})^{\alpha}$,
with an index $\alpha = -0.75\pm0.03$. This implies that the
differential distribution 
 $d{\cal N}/d{\cal R}$ follows a power-law distribution
of slope $\beta=\alpha -1 = -1.75$.

We suggest following Montmerle et~al.  (\cite{montmerle83}) that this power-law behavior,
may be explained in terms of variability due to stellar flares, if
interpreted in terms of stochastic relaxation phenomenon (Rosner \& Vaiana
1978), and dominating the \xr activity of the underlying stars.  Such a
power-law behavior is seen in the solar flares in radio, optical, soft and hard
\xr emission with the power-law index $-\beta$=1.1--3.0 (see review in
Aschwanden et~al.  \cite{aschwanden98}).  For soft \xr emission $-\beta$=1.7--1.9, which is
consistent with our result, and supports the analogy with the solar magnetic
activity.

\section{X-ray detectability of the embedded T~Tauri star population}
\label{properties}

\label{TTS_sample}

We use the information given both by the \ISOCAM survey and by our \HRI deep
exposure to study the X-ray detected TTS population of the \ro Oph dense
cores.  We restrict the following studies to the {\it HRI}/\ISOCAM overlapping
area.\footnote{We took $19.2\arcmin$ for the \HRI field radius, because one of
our sources (ROXRF31=SR9), is detected up to this angle from the axis in the
Core~F field.}  This area comprises 98 Class~II sources (classified from
ground-based and \ISO observations), including 52 new \ISOCAM Class~II sources,\footnote{We have 
excluded three new \ISOCAM Class~II sources for which we have
only the $K$ magnitude, and thus no A$_{\mathrm{V}}$ estimate.  L$_{\star}$ for
these sources must be small, and/or A$_{\mathrm{V}}$ high, which implies a high
upper limit on the intrinsic L$_{\mathrm{X}}$.  This does not affect the
statistical results.}  and 35 Class~III sources (characterized as YSO from \xr
or radio observations, and classified as Class~III sources from ground-based and
\ISO observations), including 21 new Class~III sources (\HRI or \PSPC \xr
sources without IR excess observed by {\it ISOCAM}).  We will call these sources
the ``TTS sample''.

Our \HRI observation detected a large number of sources, yet these 
constitute only 30$\%$ of the ``TTS sample''. In this section, we examine the 
reasons why the other TTS were not detected, and in particular whether the 
undetected TTS form a separate population of genuinely X-ray weak objects.

	\subsection{X-ray vs. stellar luminosities}
	\label{Lstar-Lx}

First, to know more about the X-ray properties of the members of the whole TTS 
sample (with upper limits if they are not detected with the {\it HRI}), we examine 
whether a correlation exists between the X-ray luminosity and the stellar 
luminosity (both corrected from extinction), and if so, whether it is the same 
as the one found in $\rho$~Oph by CMFA.
We chose for each Core~F \xr source its lowest \xr luminosity (including \HRI upper limits) 
to minimize the effects of \xr variability.  
For TTS undetected by the {\it HRI}, we estimate count rate upper limits 
($3.25\,\sigma$)\footnote{When the YSO is in both Core~A and Core~F field,
we use the longer Core~F field exposure to estimate the count rate upper limit.
using the EXSAS command COMPUTE/UPPER\_LIMITS, 
and we use the extinction estimate from Bontemps et~al. (\cite{bontemps00}) 
to compute the corresponding limit on the intrinsic X-ray luminosity.} 

\begin{figure}[!t]
\resizebox{\hsize}{!}{\includegraphics{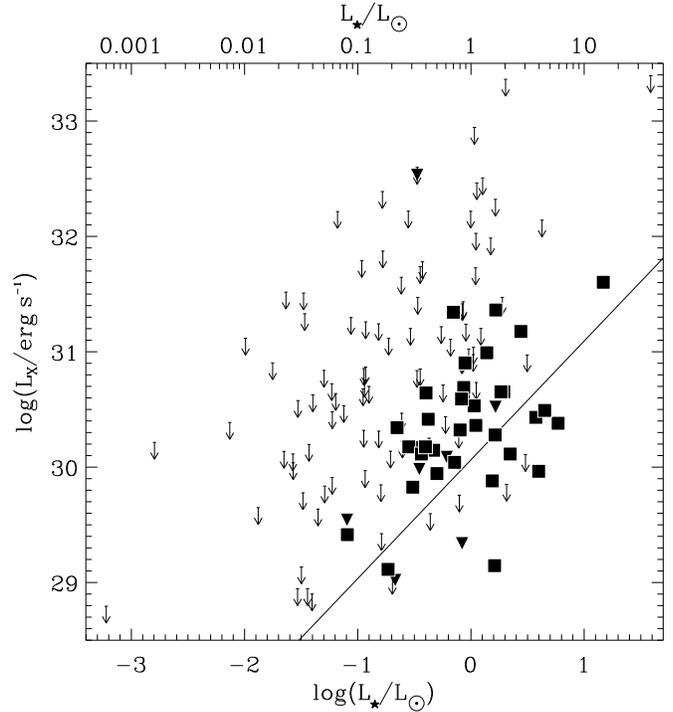}}
\caption{Intrinsic \xr and stellar luminosity correlation for the ``T~Tauri star sample''.
The squares represent the T~Tauri stars detected with the {\it HRI} in Core~A observation,
and the T~Tauri stars detected with the {\it HRI} in Core~F observations 
for which the lowest \xr luminosity of the 3 observations is not an upper limit.  
Downward triangles correspond to the T~Tauri stars detected with the {\it HRI} in Core~F observations 
for which the lowest \xr luminosity of the 3 observations is an upper limit.
Arrows are the hundred upper limits for the \HRI undetected T~Tauri stars.
The highest upper limits correspond to high extinction sources 
(see Fig.~\ref{detections}).
The solid line shows the correlation found using {\it ASURV}:
log($L_{\mathrm X}$/erg\,s$^{-1}$) = (1.0$\pm$0.2) $\times$ log($L_{\star}/L_{\odot}$) + 30.1.}
\label{Lx-Lstar-correlation}
\end{figure}

To establish the existence of a linear correlation between
$\log{L_{\mathrm X}}$ and $\log{L_{\star}}$, we performed three statistical 
tests using ASURV: Cox's
proportional hazard model, the generalized Kendall $\tau$ test, and Spearman's
\ro test.  The probability of the null hypothesis (i.e. that this correlation is
not present) is $<10^{-4}$ for each of the three tests.  Thus, a strong linear
correlation between $\log{L_{\mathrm{X}}}$ and $\log{L_{\star}}$ is indeed present.  
We found the linear regression coefficients by using the Estimation Maximization (EM) 
algorithm under Gaussian assumptions and the Buckley-James method.  
The Buckley-James method gave results similar to those of the EM algorithm, 
but with a larger uncertainty on the slope.
As the Buckley-James method is semi-nonparametric, this suggests
that the residuals of the linear correlation may be non-Gaussian.  
We thus conservatively keep the slope uncertainty given by the 
Buckley-James method.
The $L_{X}$--$L_{\star}$ correlation is then given by (see Fig.\ref{Lx-Lstar-correlation}):  
log($L_{\mathrm X}$/erg\,s$^{-1}$) = (1.0$\pm$0.2) $\times$ log($L_{\star}/L_{\odot}$) + 30.1.  
We note that the censoring fraction is so high that the correlation line misses most of the data
points and depends entirely on the location of the few lowest detections.
The correlation dispersion may be due to \xr variability, 
and also to TTS spectral type and age differences:  
Neuh{\"a}user et~al. (\cite{neuhaeuser95}) points out that the ratio $L_{\mathrm X}/L_{\star}$ increases with decreasing
effective temperature, and shows a variation of L$_{\mathrm X}$ with age.

This correlation spans three orders of magnitude in L$_{\mathrm{X}}$ and
two in L$_\star$. 
The slope of this correlation, $a$, is equal to 1.0, and the TTS \xr luminosity 
is then approximatively given by the simple proportionality:  
$L_{\mathrm X}/L_{\star} \sim 10^{-4}$.  
We thus confirm that the characteristic  for TTS
in \ro~Oph is 10$^{-4}$, with a large dispersion up to a level $\sim 10^{-3}$. 
There is no evidence for the ``saturation'' effect seen
at this level in late type main sequence stars, and attributed to the
complete filling of the stellar surface by active regions (Fleming et~al. 1989).

A similar correlation between L$_{\mathrm X}$ and L$_{\star}$ was
found for Class~II and Class~III sources in the previous \ro~Oph study of CMFA 
(the method used to estimate L$_{\star}$ was different, 
but very similar numerically; see $\S$\ref{Cloud_extinctions}), 
but also in other star-forming regions:
Chamaeleon (Feigelson et~al. \cite{feigelson93}; Lawson et~al. \cite{lawson96}), 
Taurus-Auriga (Neuh{\"a}user et~al. \cite{neuhaeuser95}), and IC 348 (Preibisch et~al. \cite{preibisch96}).
However, the slopes may not be identical:  while Feigelson et~al. (\cite{feigelson93}), 
Preibisch et~al. (\cite{preibisch96}), and CMFA find the same $a = 1$ slope as above. 
On the other hand,
Lawson et~al. (\cite{lawson96}) found $a = 0.55$, on a better characterized, 
enlarged \xr source sample in Chamaeleon, stessing the importance of having a 
sample as complete as possible.
Nevertheless, the fact that we find the same slope as CMFA  
with an enlarged sample of \xr sources
in the same cloud is certainly a good internal consistency
check between the \PSPC and the {\it HRI}.

We find that the majority of the \xr luminosity upper limits are above the
$L_{\mathrm X}$--$L_{\star}$ correlation.  Only 5$\%$ of the X-ray undetected
TTS are below the correlation, mixed with X-ray detected TTS.  This is
consistent with the idea that {\it all} TTS in \ro~Oph may be \xr emitters
with $L_{\mathrm X}/L_{\star} \sim 10^{-4}$.  Therefore, the TTS
undetected by the \HRI do not make up a separate population, but must have X-ray
properties comparable to that of the detected population, verifying the same
correlation.

	\subsection{The X-ray undetected T~Tauri star population}
	\label{X_versus_IR}

\begin{figure}[!t] 
\resizebox{\hsize}{!}{\includegraphics{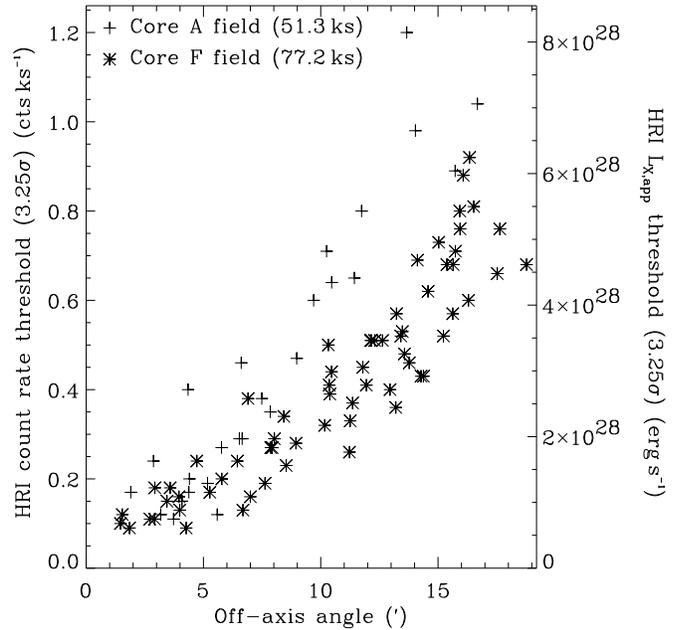}}
\caption{\HRI count rate thresholds (3.25\,$\sigma$) 
vs. off-axis distance ({\it bottom scale}).
{\it Left (resp. right) scale:} \HRI count rate (resp. apparent \xr luminosity) threshold (3.25\,$\sigma$).
Plus signs (asterisks) represent Core~A (F) T~Tauri stars of the ``T~Tauri star sample'' undetected 
with the {\it HRI}.}
\label{upper} 
\end{figure}

Using the previous correlation between the stellar and \xr luminosities, we
can now estimate for each member of the TTS sample the intrinsic \xr
luminosity in the \ROSAT energy band,\footnote{We note that this method 
attributes a larger \xr luminosity to the TTS having X-ray luminosities below the 
$L_{\mathrm X}$--$L_{\star}$ correlation. By this effect X-ray undetected TTS can be 
put above the \HRI detection threshold, but this concerns only 5 cases, see Fig.~\ref{upper}.} 
and compare it with the \HRI detection threshold to understand the X-ray detectability 
of the TTS sample with the {\it HRI}. 
However, the comparison is not straightforward, since the \HRI detection
threshold depends on both instrumental effects and extinction along the line
of sight.

Fig.~\ref{upper} shows the instrumental effects:  the \HRI count rate
threshold ($3.25\,\sigma$) increases away from the pointing axis.  We interpret this
dependence as the consequence of the point spread function degradation and
reduced sensitivity off-axis of the \ROSAT mirrors (David et~al.  \cite{david97}).
With the X-ray spectrum assumptions described in
\S\ref{derivation_from_source_counts}, we have determined using EXSAS the
conversion factor, $f$, between the \HRI counts and the apparent \xr
luminosity (i.e., in the absence of extinction) in the \ROSAT energy band
(0.1--2.4\,keV), $L_{\mathrm{X,app}}$.  We find:  $f=6.8 \times
10^{28}$\,erg\,cts$^{-1}$\,s for $d=140$\,pc.  The minimum X-ray
luminosity for a $3.25\,\sigma$ \HRI detection, $L_{\mathrm{X,min}}$, ranges
from $\sim 7 \times 10^{27}$\,erg\,s$^{-1}$ on-axis (angle=0\arcmin) to
$\sim 7 \times 10^{28}$\,erg\,s$^{-1}$ off-axis (angle=19.2\arcmin) (see
Fig.~\ref{upper}).

In case the X-ray sources suffer some extinction equivalent to A$_\mathrm V$
magnitudes, the values of $L_{\mathrm{X,min}}$ on-axis and off-axis must be
corrected to obtain the corresponding {\it intrinsic} minimum X-ray
luminosities as a function of A$_\mathrm V$:  if a source is heavily extincted, this
minimum may be up to two orders of magnitude higher or more than in the
absence of extinction (see for example the high values of the upper limits of Fig.~\ref{Lx-Lstar-correlation}).

\begin{figure*}[p]
\resizebox{\hsize}{!}{\includegraphics{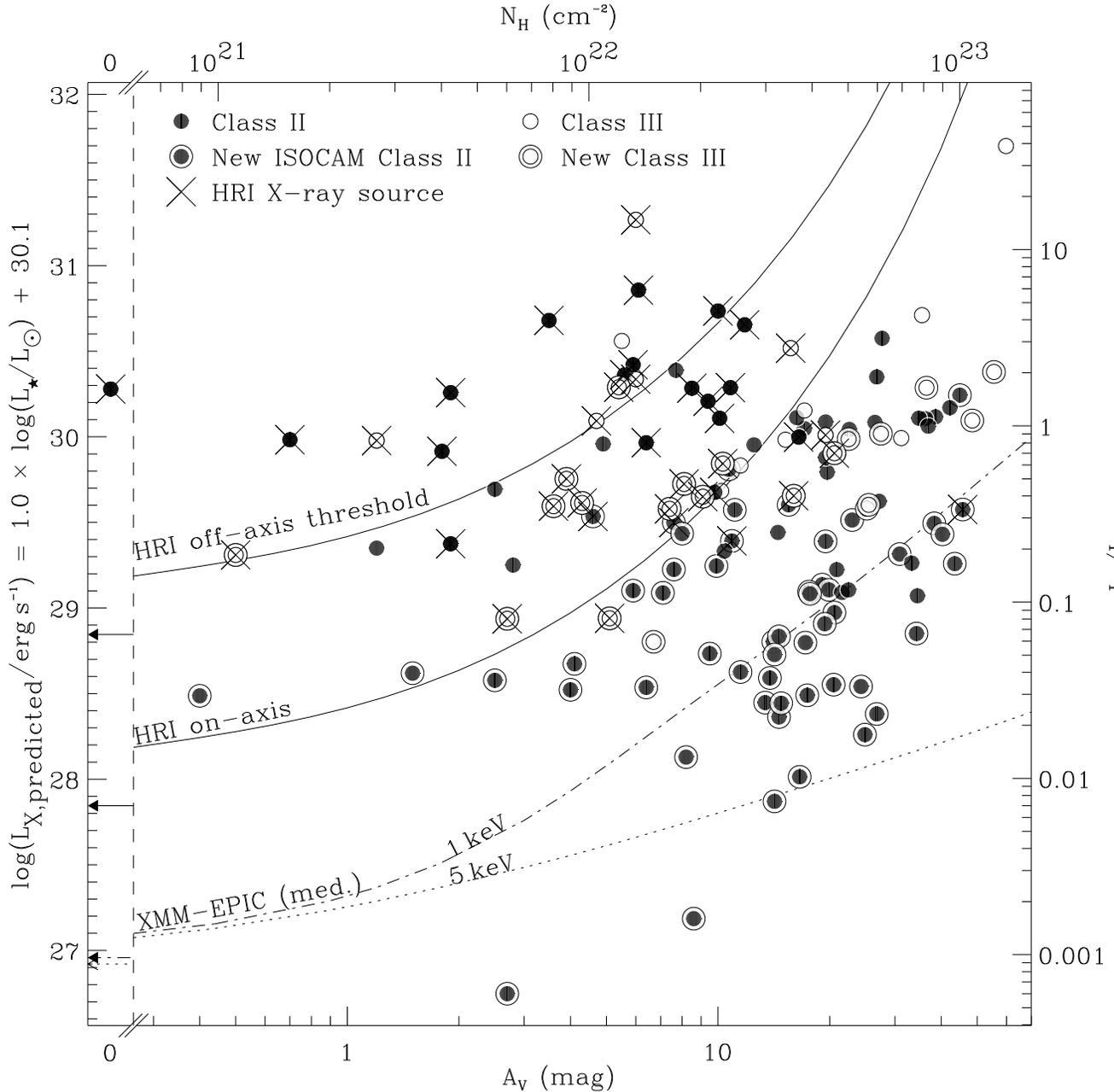}}
\caption{Detectability of the ``T~Tauri star sample'' with the \HRI vs. extinction.
{\it Right scale:} stellar luminosity determined from near-IR photometry; 
{\it left scale:} \xr intrinsic luminosity predicted from the $L_{\mathrm{X}}$--$L_{\star}$ correlation 
(see Fig.~\ref{Lx-Lstar-correlation}); 
{\it top scale:} extinction expressed in N$_{\mathrm H}$;
{\it bottom scale:} extinction expressed in A$_{\mathrm{V}}$. 
``New'' \ISOCAM Class~II sources are IR sources with IR excess discovered by {\it ISOCAM}. 
``New'' Class~III sources are IR sources characterized as YSO candidates from \xr or radio observations, 
and for which \ISOCAM observed no IR excess.
Crosses indicate T~Tauri stars detected with the {\it HRI}. 
Solid curves show the detection threshold for the \ROSAT \HRI (on- and off-axis), assuming 
a Raymond-Smith spectrum with $kT_{\mathrm{X}}$=1\,keV and an exposure time of 77\,182\,s.
Several T~Tauri stars were detected despite being below the nominal \HRI detection threshold:
these sources have likely been detected in a flaring state (see text for details).
Dashed-dotted curves show the {\it XMM-EPIC} (medium filter) detection threshold for 
the same spectrum and exposure time; we assumed a noise of $4 \times 10^{-4}$\,cts\,s$^{-1}$ for the 
{\it EPIC-pn}, and deduced the detection threshold for the all {\it EPIC} instrument applying
a 0.75 factor (see the {\it XMM} Users' Handbook; Dahlem et~al. 1999).
The dotted curves show the {\it XMM-EPIC} (medium filter) detection threshold for an high 
plasma temperature of 5\,keV (flare).
The \xr detection of the ``T~Tauri star sample'' is limited by the \HRI sensitivity.
{\it XMM-EPIC} thanks to its increased sensitivity and its energy range (0.2--12\,keV),
is less sensitive to extinction, and should be able to detect 
numerous new Class~III sources, as well as the low-luminosity Class~II
sources discovered by \ISOCAM} (see text for details).
\label{detections} 
\end{figure*}

Fig.~\ref{detections} plots the X-ray luminosities of the TTS sample as a function of $A_\mathrm V$
(or $N_\mathrm H$).  These X-ray luminosities were estimated from the stellar luminosities using
the correlation discussed in the previous section.  
The points are compared with the {\it HRI} threshold
curves $L_{\mathrm{X,min}} = f(A_\mathrm V)$, computed both on-axis and off-axis.
The \HRI detected TTS (crossed dots) are found to be rather bright ($\sim 10^{29} - \sim
10^{31}$\,erg\,s$^{-1}$) and weakly extincted ($A_{\mathrm{V}} \le
10$).  The undetected TTS have estimated X-ray luminosities below the
computed \HRI detection threshold.  In particular, we understand why the new
\ISOCAM Class~II sources (Bontemps et~al.  \cite{bontemps00}), characterized both by low
stellar luminosities ($\sim 0.05$ L$_\odot$) (and thus presumably low
predicted \xr luminosities, $\sim 6 \times 10^{28}$\,erg\,s$^{-1}$), and
relatively high extinctions ($A_{\mathrm{V}} \sim 20$), could not have been
detected with our \HRI observation.{\footnote{A few faint sources were
however detected in spite of being below the nominal \HRI detection
threshold:  they were probably in an X-ray flaring state at the time of the
observations.  In particular \object{ROXRF32} = \object{GY238}, far below the \HRI
detection threshold, was detected only in the third Core~F exposure, which
supports this interpretation.}}

Now {\it ISOCAM} cannot {\it per se} recognize Class~III YSO among its
sources without IR excess, but X-rays can.  However, a reliable census of
Class~III sources in $\rho$~Oph is {\it de facto} limited by the sensitivity of
\xr observations:  Fig.~\ref{detections} shows that the number of detected
Class~III sources decreases for low L$_{\mathrm X}$ and high A$_{\mathrm V}$;
roughly speaking Class~III sources are mainly detected above L$_\mathrm{X,min}
\sim 10^{29.6}$\,erg\,s$^{-1}$ (or equivalently
L$_\mathrm{\star,min}$=0.35\,L$_\odot$), and below A$_\mathrm{V,min} \sim 30$.

This strongly suggests that unknown Class~III sources may exist.  We can
estimate their number at least in regions at the periphery of cloud cores, by
using the fact that the WTTS/CTTS ratio (or equivalently the Class~III/Class~II
source ratio) is $\sim 1$, and also that the \HRI is equally sensitive to
Class~III and Class~II sources (see \S4).  In the {\it HRI/ISOCAM} overlapping area, this
ratio is 19/22 $\sim 1$; on a comparable area Mart{\'\i}n et al.  (\cite{martin98}) also
found a WTTS/CTTS ratio $\sim 1$.  Since the ``TTS sample'' comprises 88
Class~II sources and 35 Class~III sources above L$_\star \sim 0.03$\,L$_\odot$, 
we predict that $ (88\times19/22)-35 \sim 40$ Class~III sources
remain to be discovered in X-rays in the {\it HRI/ISOCAM} overlapping area 
above L$_\mathrm{X} \sim 3 \times 10^{28}$\,erg\,s$^{-1}$.  
These sources are not seen now either ($i$) because they are too
faint in X-rays ($L_X \leq L_\mathrm{X,min}$) --or equivalently from the existence of an
$L_\mathrm{X}~vs.~L_\star$ proportionality, too faint in stellar luminosity ($L_\star
\leq L_{\star,min}$)-- or ($ii$) too absorbed ($A_V \geq A_\mathrm{V,min}$), or a
combination of both.  {\it XMM-Newton} will be an ideal tool to reveal such a
large number of unknown Class~III sources in the future (see $\S$\ref{Summary}),
but, as shown in the next section, we can already figure out their nature to a
large extent.

\section{The unknown Class~III source population}
\label{origins}

In this section, we seek to characterize the suspected unknown 
Class~III source population, which is likely to exist on the basis of the  
Class~II source findings by \ISOCAM and other IR observations. We will 
use (i) the spatial distribution of all known Class~II sources, and (ii) 
the extinction map derived from C$^{18}$O observations.

	\subsection{Compared spatial distributions of the Class~II 
and Class~III sources}

The conventional wisdom is that Class~III sources are descendants of 
Class~II sources after dispersion of their disks (e.g., Lada 1987; AM). As cloud 
cores have an internal velocity dispersion, stars form with an initial 
mean velocity distribution, implying that they drift away from their 
formation site (e.g., Feigelson \cite{feigelson96}). This is the usual explanation for 
the increase with distance from cloud cores of the Class~III/Class~II source 
ratio (or equivalently at that stage the WTTS/CTTS ratio), which is 
$\simlt 1$ within the core region (e.g., CMFA), and 
reaches values $\gg 1$ far from the cores (e.g., Mart{\'\i}n et~al. \cite{martin98}). 
This implies a larger spread of the spatial distribution of the Class~III source 
population compared to that of the Class~II source population.

\begin{figure}[!t] 
\resizebox{\hsize}{!}{\includegraphics{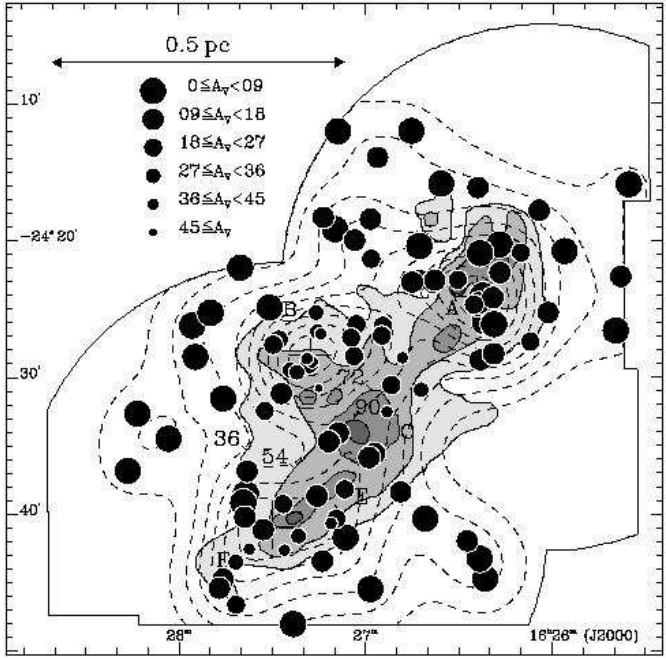}}
\caption{Spatial distribution of Class~II sources in the \ro Ophiuchi Cloud.
Black dots show the position of Class~II sources.
The dot size gives information on the source visual extinction.
The dashed contour map is an estimate of the local Class~II source surface 
density (in linear arbitray units) obtained using a sum of Gaussians 
(FWHM=6\arcmin) centered on each Class~II source position; crosses show the maxima of these peaks.
The letters show the location of DCO$^+$(J=2-1) dense cores A, B, C, E, F (see Fig.~\ref{Sources}).
The contour map shows the visual extinction (A$_\mathrm{V}$=36, 54, 72, 90) derived from C$^{18}$O column density 
(Wilking \& Lada \cite{wilking83}).
A scale of 0.5\,pc is shown for $d=140$\,pc.}
\label{Class_II_density}
\end{figure}

Let us first study the spatial distribution of the Class~II source population 
within the {\it HRI}/\ISOCAM area. We analyze the source surface density by 
using a 2-D Gaussian filter of a given FWHM
on source position. The choice of the FWHM is optimized to enhance the contrast 
between regions of low and high source density, and thus reveal any 
clustering. Fig.~\ref{Class_II_density} shows the resulting density map, in the form of 
dashed contours obtained with FWHM=6$\arcmin$. 
The Class~II sources show three strong density peaks well centered on DCO$^+$ 
cores A, B and F, which is consistent with the idea 
that most of these sources were born in these cores.  
However, in spite of its comparable DCO$^+$ line-of-sight density, 
core C appears much poorer in Class~II sources; the weaker star-forming activity of this 
core is confirmed by the presence of only one Class~I source (see Bontemps et~al. \cite{bontemps00}).

One can go one step further by comparing the source distribution with 
the matter distribution along the line-of-sight, i.e., with the extinction map. 
The DCO$^+$ radical is a good indicator of large densities 
($n \sim 10^5$--$10^7$\,cm${^{-3}}$) in cold cores, 
but the relevant regions occupy a relatively small volume; by contrast, 
C$^{18}$O, which is generally optically thin and sensitive to smaller 
densities, is a good column-density tracer. 
Using this molecule, Wilking \& Lada (\cite{wilking83}) derived an extinction map of the 
$\rho$~Oph cloud center, showing that the denser regions have an 
equivalent visual extinction A$_\mathrm{V}$ between $\sim 30$ and $\sim 100$.

Fig.~\ref{Class_II_density} displays the C$^{18}$O contours, labeled in A$_\mathrm{V}$ by steps of 
A$_\mathrm{V} \approx 20$, starting at A$_\mathrm{V} = 36$, from Wilking \& Lada (\cite{wilking83}).
Correspondingly, the Class~II sources are represented by black dots of size
decreasing with A$_\mathrm{V}$, from low extinctions (A$_\mathrm{V} < 9$) to high extinctions
(A$_\mathrm{V} \geq 45$), by steps of A$_\mathrm{V} \approx 10$.  A large majority of these
sources are seen to have moderate extinctions (A$_\mathrm{V} < 18$), 
even in the areas overlapping regions of high extinctions traced by C$^{18}$O.  
This implies that such sources are actually only moderately embedded in the cloud, 
{\it in front} of the densest regions traced by C$^{18}$O and DCO$^+$, rather than within them.  
The spatial distribution of the Class~II sources can thus be more appropriately described as
gaussian-like three-dimensional overlapping ``haloes'' around the DCO$^+$ cores
A, B and F. This also implies that at least a fraction of the Class~II sources
with high extinctions are not necessarily really embedded in the densest regions, 
but may be part of these haloes {\it behind} the dense cores.

\begin{figure}[!t] 
\resizebox{\hsize}{!}{\includegraphics{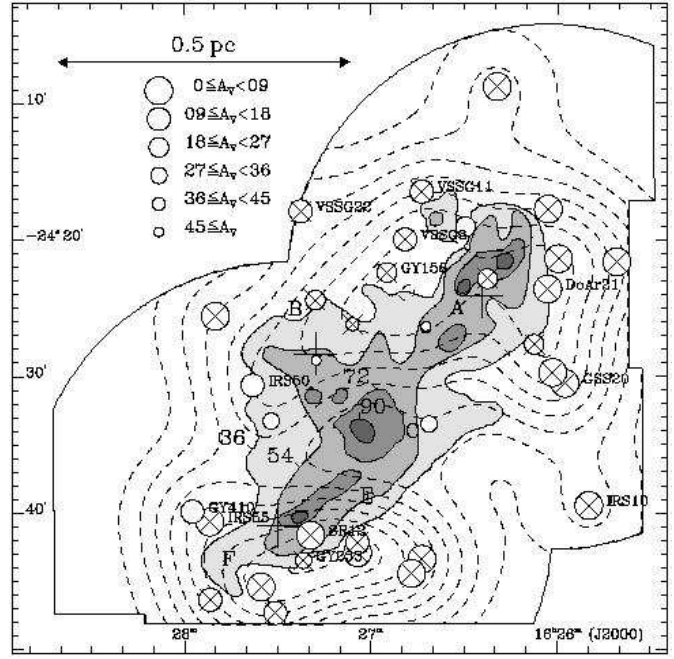}}
\caption{Spatial distribution of Class~III sources in the \ro Ophiuchi Cloud.
White dots show the position of Class~III sources excluding the early-type
star S1.
Crosses show \xr detected Class~III sources (Montmerle et~al. \cite{montmerle83}; CMFA; Koyama et~al. \cite{koyama94}; Kamata et~al. \cite{kamata97}; 
this article).
The dashed contour map is an estimate of the Class~III source surface density
(in arbitrary linear units) obtained using a sum of Gaussians 
(FWHM=8\arcmin) centered on each Class~III source position.
Plus signs show the positions of the density peaks of Class~II sources (see Fig.~\ref{Class_II_density}).
Class~III sources with known spectral types are labeled, and put in an H-R diagram in Fig.~\ref{ps99}.}
\label{Class_III_density}
\end{figure}

In the very same fashion, Fig.~\ref{Class_III_density} displays the distribution
of the known Class~III sources, this time using FWHM=8\arcmin.\footnote{A larger
FWHM must be used because there are fewer Class~III sources than Class~II
sources and because they are less clustered.}  This distribution is different
from that of Class~II sources in Fig~\ref{Class_II_density}.  With respect to
the DCO$^+$ cores, there is a strong density peak $\sim 5\arcmin$~SW of the
location of core F, and no peak associated with cores A and B:  in contrast with
the distribution of Class~II sources, there is a deficiency of Class~III sources
in the cloud center regions with high extinction.

The explanation for this apparent absence may be as follows.  Whether they lie
on the line-of-sight to regions of moderate extinction, or of high extinction,
moderate-extinction Class~II sources are found essentially everywhere.
Therefore {\it we also expect to have low-extinction Class~III sources
everywhere}, in a $\sim$ 1:1 proportion.  If they are not detected with the {\it
HRI}, it can thus only mean that they are too faint in X-rays, hence have a
small stellar luminosity.  In addition, as for Class~II sources, we must expect
along the line-of-sight to the densest regions of the cloud to also have
moderately embedded Class~III sources at the back of the cloud.  Their spatial
density should be roughly comparable to that of the unseen Class~III sources in
the front, i.e., yield a small absolute number given the compact size of the
C$^{18}$O cores.

There is also the possibility of having genuinely embedded (hence very young)
Class~III sources in these cores:  we have no information about the
Class~III/Class~II source ratio there, so it is impossible to estimate their number.
Should this number be large (such that Class~III/Class~II $> 1$ for instance),
this would be a problem for the earliest stages of YSO evolution; one rather
expects to have Class~III/Class~II $\ll 1$ if all stars form with a disk taking
at least $\simgt 10^5$\,yr to dissipate.  However, a disk stage is perhaps not
necessary for very low-mass stars, which would increase the number of very young
Class~III sources.

\subsection{Constraints on the nature of the unknown Class~III sources}

\begin{figure}[!t] 
\resizebox{\hsize}{!}{\includegraphics{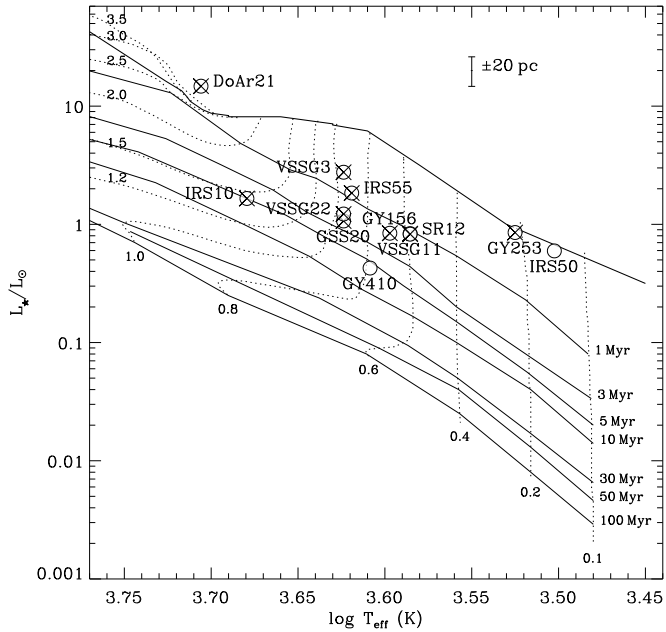}}
\caption{H-R diagram for the Class~III source with a known spectral types.
Dotted lines show Palla \& Stahler (\cite{palla99}) pre-main sequence tracks, 
continuous lines show isochrones.
The bold continuous line represents the birthline for the star build-up accretion rate ($10^{-5}$\,M$_\odot$\,yr$^{-1}$).
White dots show the position of Class~III sources with known spectral type 
(Bouvier \& Appenzeller \cite{bouvier92}; Luhman \& Rieke \cite{luhman99}; see also
Greene \& Meyer \cite{greene95}), and using the Bontemps et~al. (\cite{bontemps00}) luminosities.
Crosses show \xr detected Class~III sources (Montmerle et~al. \cite{montmerle83}; CMFA; 
Koyama et~al. \cite{koyama94}; Kamata et~al. \cite{kamata97}; this article).
We take $d=140$\,pc. The error bar shows the systematic luminosity error corresponding to a distance error 
of $\pm20$\,pc.}
\label{ps99}
\end{figure}

Let us construct the H-R diagram of the 12 Class~III sources 
in {\it HRI}/\ISOCAM area for which we know the spectral types 
from the optical observations of Bouvier \& Appenzeller (\cite{bouvier92}), and 
from the $K$-band observations of Luhman \& Rieke (\cite{luhman99}), using the 
stellar luminosities determined by Bontemps et al. (\cite{bontemps00}).\footnote{The relative differences between these 
luminosities, and the bolometric corrected ones from spectral types are lower than $20\%$.}  
Fig.~\ref{ps99} displays the result, along with the birthline and pre-main sequence 
evolutionary tracks of Palla \& Stahler (\cite{palla99}). 
According to these evolutionary tracks, the ages of the 12 Class~III sources are 
found to be spread between $\sim 0.2$ and $\sim 5$\,Myr. 

It is reasonable to assume that the unknown Class~III sources have the same age spread.
These sources are not yet detected in X-rays 
either because their intrinsic X-ray luminosities are too low, 
and/or because they have high extinctions. 
In the first case they have stellar luminosities below 0.35\,L$_\odot$,
and the isochrones imply that $M_\star \sim 0.6$\,M$_\odot$ for the oldest ones, 
and $M_\star < 0.1$\,M$_\odot$ for the youngest ones.  
In the second case, however, the unknown embedded Class~III sources can have luminosities higher than 
$L_\mathrm{\star,min}$, i.e., so that they are not necessarily very low-mass stars. 

Unless the number of Class~III sources embedded in the densest regions is very high, 
our conclusion is that {\it the bulk of the Class~III sources which are 
undetected by the \HRI and unrecognized by \ISOCAM should be made of very low-mass stars}.

\section{Summary and conclusions}
\label{Summary}

\subsection{Main observational results}

We have obtained two deep exposures of the \ro Oph cloud core region 
(d=140\,pc) with the \ROSAT {\it High Resolution Imager} (core~A:  51\,ks,
core~F:  77\,ks, in three partial exposures).  The improved position accuracy
(1$\arcsec$--6$\arcsec$) with respect to previous recent \xr observations
(\ROSAT {\it PSPC}, Casanova et~al.  \cite{CMFA}; and {\it ASCA}, Koyama et~al.  \cite{koyama94}
and Kamata et~al.  \cite{kamata97}) have allowed us to remove a number of positional
ambiguities for the detected sources.  We have cross-correlated the X-ray
positions with IR sources found in the {\it ISOCAM} survey of the same
region at 6.7 and 14.3\,$\mu$m, in addition to sources known in the optical, IR,
and radio from ground-based observations.  We thus have now at our disposal the
best-studied sample of X-ray emitting YSO in a star-forming region.  We first
summarize the main observational results of this article.

\begin{list}{}{\itemsep 0pt}

\item[(1)] We detect 63 \HRI \xr sources, and 55 are identified.
Of the 55 identified \xr sources 40 are \PSPC sources, and 9 are \ASCA sources.

\item[(2)] The IR classification (ground-based and \ISOCAM survey) for the 55
identified \xr sources yields:  one Class~I protostar (\object{YLW15}=\object{IRS43}); 23 Class~II
sources, including 4 new \ISOCAM Class~II sources; 21 Class~III sources,
including 13 new Class~III sources; 8 new Class~II or Class~III source
candidates; one early-type Class~III source (the young magnetic B3 star \object{S1}), and
one field star (the F2V star \object{HD148352}). The contamination of the sample of new
X-ray sources by field stars is negligible.

\item[(3)] There is no statistically significant difference between the X-ray
luminosity functions of {\it HRI}-detected Class~II and Class~III sources, i.e.
T~Tauri stars with and without disks, confirming that the contribution of these
disks to X-ray absorption, or emission (for instance by magnetic
reconnection between the star and the disk), must be small.

\item[(4)] X-ray variability of {\it HRI}-detected T~Tauri stars has been
studied by comparing the \HRI data with the previously obtained \PSPC data, 
and using \HRI observations done at three different epochs.  The resulting
statistics show that most of the sources are variable, and that their \xr
variability is consistent with a solar-like (hence magnetic) flare origin.

\item[(5)] We use the information given both by the \ISOCAM survey and by our
\HRI deep exposure to study the T~Tauri star population of the \ro Oph dense
cores.  We confirm that essentially all Class~II and Class~III YSO are X-ray
emitters, and that a strong correlation ($\log(L_{\mathrm X}/erg\,s^{-1}) =
(1.0\pm0.2) \times \log(L_{\star}/L_{\odot}) + 30.1$) 
exists between the X-ray luminosity and the stellar luminosity of T~Tauri stars, 
likely down to low luminosities (L$_\star \sim 0.1$\,L$_\odot$).  We confirm that the characteristic
$L_{\mathrm X}/L_{\star}$ for T~Tauri stars is $\sim$10$^{-4}$ in the $\rho$~Oph
cloud, albeit with a large dispersion.  There is no evidence for a magnetic
``saturation'' seen at a level of 10$^{-3}$ in late-type main sequence stars.

\item[(6)] However, most of the new \ISOCAM Class~II sources are not detected by
the \HRI.  We show that this is consistent with their intrinsic X-ray
luminosities being too faint if ``predicted'' using the above $L_{\mathrm X}$--$L_{\star}$ correlation.

\end{list}

\subsection{What have we learned ?}

\begin{list}{}{\itemsep 0pt}

\item[(1)] The first general conclusion we can draw from the \HRI results
presented above is a complete confirmation of the \PSPC results obtained by
CMFA.  This was not {\it a priori} obvious, since the CMFA population (\PSPC and
near-IR) overlaps, but is different from, the {\it HRI}/near-IR/\ISOCAM population
presented in this paper:  many \PSPC sources are not detected by the \HRI (see
Appendix C), and some \HRI sources are Class~II and Class~III newly classified
thanks to a combined identification with {\it ISOCAM}.  This shows that the
$L_\mathrm{X}$--$L_\star$ correlation is robust for the $\rho$~Oph TTS.

\item[(2)] The second, and perhaps most important, conclusion is the probable
existence of $\sim 40$ unknown X-ray YSO down to a limit of L$_\mathrm{X} \sim 3 \times
10^{28}$\,erg\,s$^{-1}$ in the {\it HRI/ISOCAM} overlapping area, which
should be mainly low- to very low-mass ($<0.1$--0.6\,M$_\odot$) diskless,
``Class~III TTS''.  This prediction is based both on the use of the
$L_\mathrm{X}$--$L_\star$ correlation, legitimated by its robustness, and on the discovery
of a large number of faint new IR sources by {\it ISOCAM}.  As shown below, it may be
soon verified by the next generation of X-ray satellites, namely {\it XMM-Newton} and
{\it Chandra}.  In this respect, the present paper can be taken as a ``transition''
paper between two generations of X-ray satellites.

\end{list}

Why is the detection of these ``unknown TTS'' important ?  Because they are
diskless, they are unlikely to be recognized as YSO by IR observations alone;
and because they are likely to be as numerous as the YSO with IR excess, they
have to be included in any reliable census of YSO, with an impact on such basic
quantities as the initial mass function, or the star formation efficiency,
especially if considered from an evolutionary point of view.  For instance, from
the results in this paper it is impossible to study the real connection between
the distributions of the Class~II and Class~III sources in the densest regions,
in particular to see whether the distribution of the Class~III sources is also
centered on the same DCO$^+$ cores as the Class~II sources.  The number of
Class~III sources embedded in the densest regions may, or may not, be comparable
to that of the Class~II sources, depending on the timescale for disk dispersal,
especially among low-mass YSO.  An X-ray improved census of Class~III sources
may also be crucial in determining whether a burst of star formation is
presently going on in the $\rho$~Oph cores, as some recent indications suggest
(see Mart{\'\i}n et al.  1998).  It will also allow to study the $L_\mathrm{X}$--$L_\star$ 
correlation for Class~III and Class~II sources seperately, which
was not possible in this paper (\S\label{Lstar-Lx}) due to insufficient statistics.

\subsection{The potential of {\it XMM-Newton} and {\it Chandra}}

To quantify the prospects for improvement in the \xr domain, we have computed
the detection threshold for the \xr camera {\it EPIC} aboard {\it XMM-Newton},
which was successfully launched in December 1999 (see
Fig.~\ref{detections}).  The improved sensitivity and enlarged energy range
(0.5--12\,keV) of {\it EPIC} will allow to detect the weak \ISOCAM Class~II
sources, and also to discover numerous unknown faint or embedded
Class~III sources, in particular if they have high plasma temperatures (several
keV) reached during flares, and extend the census of this population towards the
low-mass end.  In the best case, the {\it XMM-Newton} sensitivity will reach L$_\mathrm{X} \sim
10^{28}$\,erg\,s$^{-1}$ for A$_\mathrm{V} \simgt 20$, for long exposures ($>$75\,ksec).
This is nearly two orders of magnitude more sensitive than {\it ROSAT}.  In case the faint
Class~III sources turn out to be so crowed that confusion problems arise, the
excellent angular resolution of {\it Chandra} will be critical.

In \ro Oph, there are already several identified {\it bona fide} and candidate
brown dwarfs (see review in Neuh{\"a}user et~al.  \cite{neuhaeuser99}, and references therein),
and four of them have been recently detected in \xr using the \ROSAT \PSPC
archive (Neuh{\"a}user et~al.  \cite{neuhaeuser99}).  Neuh{\"a}user et~al.  have also shown that
brown dwarfs could be \xr emitters with the same ratio $\log(L_{\mathrm
X}/L_\star) \sim -4$ than for T~Tauri stars.  Thus {\it Chandra} and {\it
XMM-Newton} should be able to detect many more of these objects with low stellar
luminosity and masses, shedding a new light on their nature and early
evolution.


{\acknowledgements
We thank Francesco Palla for fruitful discussions during the 5$^{th}$ French-Italian meeting
in the island of Ponza, and the referee Fred Walter for his useful remarks.
NG is supported by the European Union (Marie Curie Individual grant; HPMF-CT-1999-00228).
EDF is partially supported by NASA contract NAS8-38252. 
We used SIMBAD maintained by the CDS (Strasbourg Observatory, France). 
We also used photographic data obtained using The UK Schmidt Telescope: original plate material 
is copyright $\copyright$ of the ROE and the AAO.}


\appendix

\section{HRI X-ray source detection}
\label{appendix:source_detection}

\begin{figure*}[ht] 
\resizebox{\hsize}{!}{\includegraphics{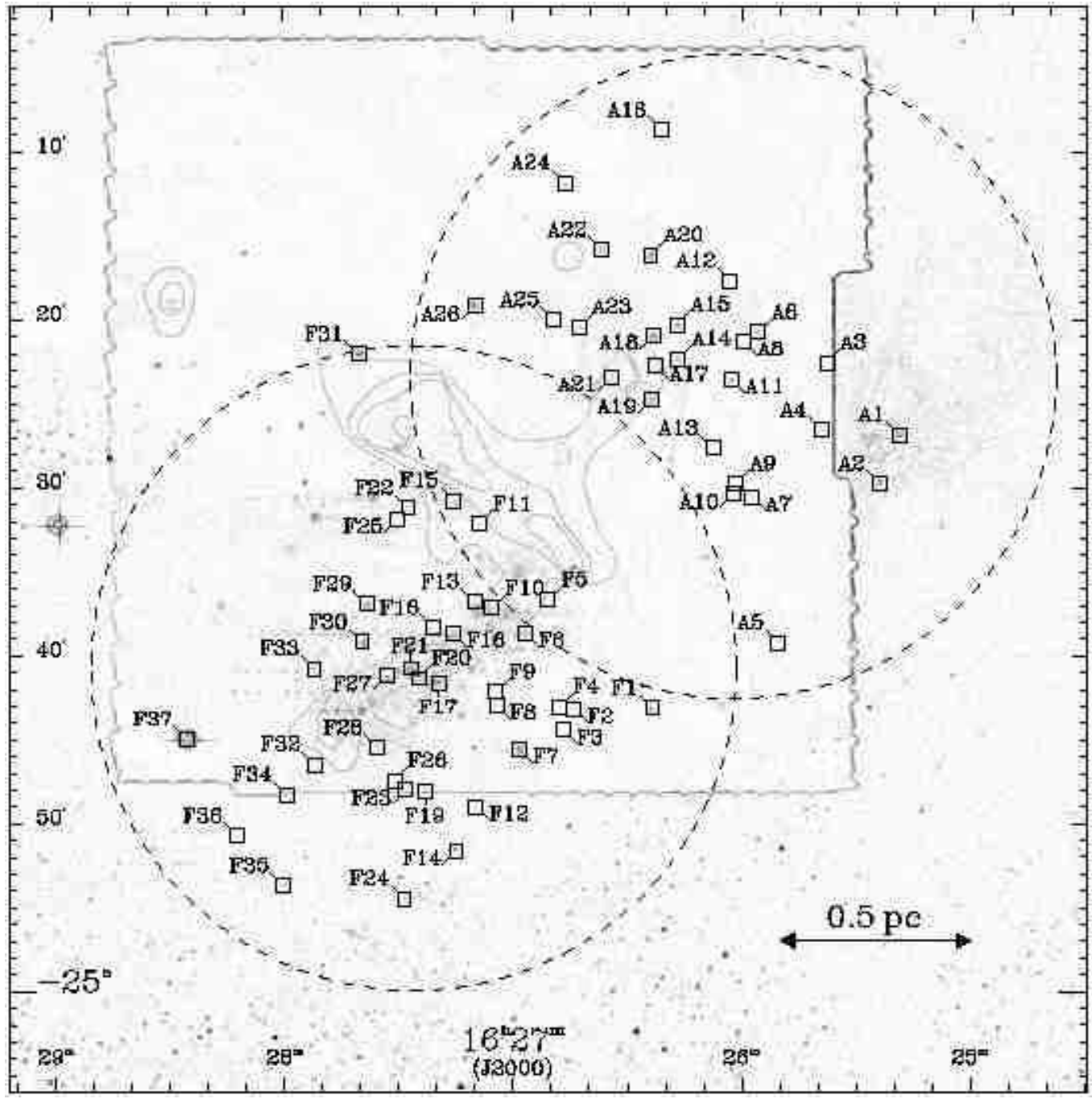}}
\caption{Sources designations from Table~\ref{tab:sources}--\ref{tab:ident}. 
The image of Fig.~\ref{Sources} was smoothed using an edge detection algorithm.}
\label{Names}
\end{figure*}

Source detection was done using EXSAS (Zimmermann et~al. \cite{zimmermann97}), and the standard 
command DETECT/SOURCES.  This command generates a local source detection by a
sliding-window technique followed by a maximum likelihood test which compares
the observed count distribution on the full resolution image (pixel of 0.5$\arcsec$)
to a model of the point spread function (PSF) and the local background 
(Cruddace et~al. \cite{cruddace88}).  
The ``likelihood of existence'' is defined as ${\cal L} = -\ln
P_0$, where $P_0$ the probability of the null hypothesis that the observed
distribution of counts is only due to a statistical background fluctuation;
$\cal L$ provides a maximum likelihood measure for the presence of a source
above the local background.  We take $\cal L$=6.8 as detection threshold ($P_0
\le 0.0011$; or $\ge 3.25\,\sigma$ for Gaussian statistics) as argued in CMFA.

For source detection, the \HRI report (David \cite{david97}) advises to screen out lower
and higher \HRI Pulse Height Analyzer (\ha PHA) channels, which are found to have
the highest background.  However, source counts should always be determined
using all the 1--15 channels to cancel out the uneven \HRI efficiency
distribution across the detector area (S.  D{\"o}bereiner, private communication).
We decided to search \xr sources above a fixed detection threshold in channels
1--15 and 3--8.\footnote{In the core~F image \#3, 
many 1--15 channel detections with high $\cal L$ were not found in the 3--8 channel band
Thus, we decided to take channels 2--8 instead of 3--8 in the three exposures.  
This third observation was the last of our program, one year after the Core~A 
observation (see Table~\ref{tab:log}).  
According to Prestwich et~al. (\cite{prestwich98}), the mean pulse height
decreases by 0.5\,channels\,year$^{-1}$:  this effect might explains why we must
decrease the lower channel boundary from 3 to 2.  The upper channel boundary
seems to be less sensitive, probably due to fewer counts in upper channels
(see David et~al. \cite{david97}).}  EXSAS gave us list of \xr detections with positions,
one sigma error box ($\sigma_{X}$), $\cal L$, and count rate.  We removed
sources detected in channels 1--15 but not in 3--8, considering these detections
as spurious.  For instance, there are two hot spots in the Core~A
observation in the South-East corner of the {\it HRI}.  Since hot spots are usually
considered as spurious detections, this criterion automatically removes them.

The astrometry must be corrected from offsets of typically a few arcseconds due
to the time-dependent boresight error in the \ROSAT aspect system.  To do this,
IR or optical counterparts in a 10$\arcsec$ radius circle around the
\xr sources were searched.  We then selected a sample of \xr sources with an unambiguous
counterpart and 1\,$\sigma_{X}$ (half width) error box $\leq 1\arcsec$, comparable to the
IR/optical typical position error box ($\sigma_{\star} \sim 1\arcsec$).  
Then, offsets in right ascension ($\alpha$) and declination ($\delta$) were estimated by
individual offset weighted mean:  \( \alpha_{offset}= \sum_{i=1}^{n} w_i\times
(\alpha_{\mathrm{X,i}}-\alpha_{\star,\rm i}),~\delta_{offset}= \sum_{i=1}^{n} w_{i}\times
(\delta_{\mathrm{X,i}}-\delta_{\star,\rm i})\), with \( w_i= (
1/\sigma_{\mathrm{X,i}}+1/\sigma_{\star,\rm i} )/ \sum_{i=1}^{n}
(1/\sigma_{\mathrm{X,i}}+1/\sigma_{\star,\rm i} ) \).  We found offsets ranging from -0.5$\arcsec$
to 2.5$\arcsec$.  We subtracted these offsets, and checked the quality of our
astrometry by estimating sample residuals mean before and after offsets
subtraction:  \( \sigma_{shift}= \frac{1}{n} \times \sum_{i=1}^{n}
\sqrt{(\alpha_{\mathrm{X,i}}-\alpha_{\star,\rm i}-\alpha_{offset})^2 +
(\delta_{\mathrm{X,i}}-\delta_{\star,\rm i}-\delta_{offset})^2} \).  We found
$\sigma_{shift}$ ranging from 0.5$\arcsec$ to 1.2$\arcsec$.  $\sigma_{shift}$ and
$\sigma_{\star}$ ($=1.2\arcsec$) were then quadratically added to $\sigma_{X}$ to
obtain an error box radius after astrometric correction
($\sigma_{total}^2=\sigma_{\mathrm X}^2+\sigma_{shift}^2+\sigma_{\star}^2$).  In the case
of the three different core~F observations, the images were aligned and merged
after astrometric correction to obtain a single deep \HRI exposure of 77.2\,ks.
Source detection was subsequently performed as described in the article.

Table~\ref{tab:sources} lists the \HRI \xr sources, 
for which we adopt, in Col.~1, the same acronym as in CMFA:  
``ROXR'' (for \ro Oph \xr $ROSAT$ source), 
followed by ``A'' or ``F''.\footnote{The ``legal'' designation for a new ROSAT source is RXJHHMM.$m$
$\pm$ DDMM, where J stands for coordinates in J2000 and $m$ is the cut (not
rounded) decimal value.  Our tables do not use this notation for simplicity in
the discussion, but the correct designation is easy to reconstruct
from the position given if required.  For example, ROXRF14 = ROXs20B = RXJ162714-2430.}  
Fig.~\ref{Names} indicates the source numbering.

\begin{table*}[p]
\caption{{\it HRI} X-ray sources in the $\rho$ Oph cloud cores A \& F.}
\smallskip
\scriptsize
\begin{tabular}{ccccccrrrrr}
\hline
\hline
\vspace{-.1cm}\\
     & \multicolumn{2}{c}{J2000} & \multicolumn{2}{c}{B1950}& & & \multicolumn{4}{c}{COUNT RATE} \\
       &\multicolumn{2}{c}{\hrulefill}&\multicolumn{2}{c}{\hrulefill}& &  & \multicolumn{4}{c}{\hrulefill} \vspace{.1cm}\\
 ROXR  & \multicolumn{1}{c}{$\alpha$} & \multicolumn{1}{c}{$\delta$} &
  \multicolumn{1}{c}{$\alpha$} & \multicolumn{1}{c}{$\delta$} & \multicolumn{1}{c}{$\pm$} & \multicolumn{1}{c}{$\cal L$} & \multicolumn{1}{c}{\#1} & \multicolumn{1}{c}{\#2} & \multicolumn{1}{c}{\#3} & \multicolumn{1}{c}{\#1+2+3} \vspace{.1cm}\\
      &  & & & & [\arcsec] &  & \multicolumn{1}{c}{[cts/ks]}& \multicolumn{1}{c}{[cts/ks]} & \multicolumn{1}{c}{[cts/ks]} & \multicolumn{1}{c}{[cts/ks]} \\ 
\multicolumn{1}{c}{(1)} & (2) & (3) & (4) & (5) & (6) & \multicolumn{1}{c}{(7)} & \multicolumn{1}{c}{(8)}&  \multicolumn{1}{c}{(9)} & \multicolumn{1}{c}{(10)} & \multicolumn{1}{c}{(11)}\\
\hline
\vspace{-0.25cm}\\
A1~ &  16$^{\mathrm h}$25$^{\mathrm m}$19\fs0 & -24$\degr$26$\arcmin$52$^{\prime\prime}$ &16$^{\mathrm h}$22$^{\mathrm m}$17\fs6 & -24$\degr$20$\arcmin$04$^{\prime\prime}$ & 2 & 5765.4 & 40.7$\pm$0.9\hspace{0.6cm} \\
A2~ &  16$^{\mathrm h}$25$^{\mathrm m}$24\fs2 & -24$\degr$29$\arcmin$47$^{\prime\prime}$ &16$^{\mathrm h}$22$^{\mathrm m}$22\fs7 & -24$\degr$22$\arcmin$59$^{\prime\prime}$ & 3 &   15.4 &  0.8$\pm$0.2\hspace{0.6cm} \\
A3~ &  16$^{\mathrm h}$25$^{\mathrm m}$38\fs0 & -24$\degr$22$\arcmin$35$^{\prime\prime}$ &16$^{\mathrm h}$22$^{\mathrm m}$36\fs7 & -24$\degr$15$\arcmin$47$^{\prime\prime}$ & 2 &   18.5 &  0.4$\pm$0.1\hspace{0.6cm} \\
A4~ &  16$^{\mathrm h}$25$^{\mathrm m}$39\fs4 & -24$\degr$26$\arcmin$36$^{\prime\prime}$ &16$^{\mathrm h}$22$^{\mathrm m}$38\fs0 & -24$\degr$19$\arcmin$49$^{\prime\prime}$ & 2 &   20.0 &  0.6$\pm$0.1\hspace{0.6cm} \\
A5~ &  16$^{\mathrm h}$25$^{\mathrm m}$50\fs6 & -24$\degr$39$\arcmin$16$^{\prime\prime}$ &16$^{\mathrm h}$22$^{\mathrm m}$48\fs9 & -24$\degr$32$\arcmin$29$^{\prime\prime}$ & 2 &  742.2 & 16.1$\pm$0.7\hspace{0.6cm} \\
A6~ &  16$^{\mathrm h}$25$^{\mathrm m}$56\fs0 & -24$\degr$20$\arcmin$47$^{\prime\prime}$ &16$^{\mathrm h}$22$^{\mathrm m}$54\fs7 & -24$\degr$14$\arcmin$01$^{\prime\prime}$ & 2 &  101.2 &  1.9$\pm$0.2\hspace{0.6cm} \\
A7~ &  16$^{\mathrm h}$25$^{\mathrm m}$57\fs5 & -24$\degr$30$\arcmin$34$^{\prime\prime}$ &16$^{\mathrm h}$22$^{\mathrm m}$56\fs0 & -24$\degr$23$\arcmin$48$^{\prime\prime}$ & 2 &  231.1 &  3.0$\pm$0.3\hspace{0.6cm} \\
A8~ &  16$^{\mathrm h}$25$^{\mathrm m}$59\fs6 & -24$\degr$21$\arcmin$20$^{\prime\prime}$ &16$^{\mathrm h}$22$^{\mathrm m}$58\fs3 & -24$\degr$14$\arcmin$34$^{\prime\prime}$ & 2 &   23.0 &  0.7$\pm$0.2\hspace{0.6cm} \\
A9~ &  16$^{\mathrm h}$26$^{\mathrm m}$01\fs6 & -24$\degr$29$\arcmin$47$^{\prime\prime}$ &16$^{\mathrm h}$23$^{\mathrm m}$00\fs1 & -24$\degr$23$\arcmin$01$^{\prime\prime}$ & 2 &   16.8 &  0.5$\pm$0.1\hspace{0.6cm} \\
A10 &  16$^{\mathrm h}$26$^{\mathrm m}$02\fs3 & -24$\degr$30$\arcmin$26$^{\prime\prime}$ &16$^{\mathrm h}$23$^{\mathrm m}$00\fs8 & -24$\degr$23$\arcmin$40$^{\prime\prime}$ & 3 &    8.8 &  0.4$\pm$0.1\hspace{0.6cm} \\
\hline
\vspace{-0.25cm}\\
A11 &  16$^{\mathrm h}$26$^{\mathrm m}$02\fs9 & -24$\degr$23$\arcmin$35$^{\prime\prime}$ &16$^{\mathrm h}$23$^{\mathrm m}$01\fs6 & -24$\degr$16$\arcmin$50$^{\prime\prime}$ & 2 & 3581.6 & 24.1$\pm$0.7\hspace{0.6cm} \\
A12 &  16$^{\mathrm h}$26$^{\mathrm m}$03\fs2 & -24$\degr$17$\arcmin$45$^{\prime\prime}$ &16$^{\mathrm h}$23$^{\mathrm m}$01\fs9 & -24$\degr$10$\arcmin$59$^{\prime\prime}$ & 2 &   47.4 &  1.0$\pm$0.2\hspace{0.6cm} \\
A13 &  16$^{\mathrm h}$26$^{\mathrm m}$07\fs7 & -24$\degr$27$\arcmin$41$^{\prime\prime}$ &16$^{\mathrm h}$23$^{\mathrm m}$06\fs2 & -24$\degr$20$\arcmin$55$^{\prime\prime}$ & 2 &   19.4 &  0.7$\pm$0.2\hspace{0.6cm} \\
A14 &  16$^{\mathrm h}$26$^{\mathrm m}$16\fs8 & -24$\degr$22$\arcmin$22$^{\prime\prime}$ &16$^{\mathrm h}$23$^{\mathrm m}$15\fs5 & -24$\degr$15$\arcmin$37$^{\prime\prime}$ & 2 &  169.1 &  2.5$\pm$0.2\hspace{0.6cm} \\
A15 &  16$^{\mathrm h}$26$^{\mathrm m}$17\fs1 & -24$\degr$20$\arcmin$20$^{\prime\prime}$ &16$^{\mathrm h}$23$^{\mathrm m}$15\fs7 & -24$\degr$13$\arcmin$36$^{\prime\prime}$ & 2 &  235.1 &  2.9$\pm$0.3\hspace{0.6cm} \\
A16 &  16$^{\mathrm h}$26$^{\mathrm m}$21\fs3 & -24$\degr$08$\arcmin$45$^{\prime\prime}$ &16$^{\mathrm h}$23$^{\mathrm m}$20\fs2 & -24$\degr$02$\arcmin$01$^{\prime\prime}$ & 5 &   12.5 &  1.3$\pm$0.3\hspace{0.6cm} \\
A17 &  16$^{\mathrm h}$26$^{\mathrm m}$22\fs7 & -24$\degr$22$\arcmin$49$^{\prime\prime}$ &16$^{\mathrm h}$23$^{\mathrm m}$21\fs3 & -24$\degr$16$\arcmin$04$^{\prime\prime}$ & 3 &    7.9 &  0.3$\pm$0.1\hspace{0.6cm} \\
A18 &  16$^{\mathrm h}$26$^{\mathrm m}$23\fs4 & -24$\degr$20$\arcmin$59$^{\prime\prime}$ &16$^{\mathrm h}$23$^{\mathrm m}$22\fs1 & -24$\degr$14$\arcmin$14$^{\prime\prime}$ & 2 &   95.1 &  1.4$\pm$0.2\hspace{0.6cm} \\
A19 &  16$^{\mathrm h}$26$^{\mathrm m}$23\fs9 & -24$\degr$24$\arcmin$47$^{\prime\prime}$ &16$^{\mathrm h}$23$^{\mathrm m}$22\fs5 & -24$\degr$18$\arcmin$02$^{\prime\prime}$ & 3 &   10.5 &  0.4$\pm$0.1\hspace{0.6cm} \\
A20 &  16$^{\mathrm h}$26$^{\mathrm m}$24\fs1 & -24$\degr$16$\arcmin$11$^{\prime\prime}$ &16$^{\mathrm h}$23$^{\mathrm m}$22\fs8 & -24$\degr$09$\arcmin$27$^{\prime\prime}$ & 2 &   18.9 &  0.7$\pm$0.2\hspace{0.6cm} \\
\hline
\vspace{-0.25cm}\\
A21 &  16$^{\mathrm h}$26$^{\mathrm m}$34\fs2 & -24$\degr$23$\arcmin$26$^{\prime\prime}$ &16$^{\mathrm h}$23$^{\mathrm m}$32\fs8 & -24$\degr$16$\arcmin$43$^{\prime\prime}$ & 2 &   54.7 &  1.0$\pm$0.2\hspace{0.6cm} \\
A22 &  16$^{\mathrm h}$26$^{\mathrm m}$36\fs9 & -24$\degr$15$\arcmin$53$^{\prime\prime}$ &16$^{\mathrm h}$23$^{\mathrm m}$35\fs7 & -24$\degr$09$\arcmin$10$^{\prime\prime}$ & 3 &   10.9 &  0.6$\pm$0.2\hspace{0.6cm} \\
A23 &  16$^{\mathrm h}$26$^{\mathrm m}$42\fs8 & -24$\degr$20$\arcmin$29$^{\prime\prime}$ &16$^{\mathrm h}$23$^{\mathrm m}$41\fs5 & -24$\degr$13$\arcmin$46$^{\prime\prime}$ & 3 &   13.1 &  0.6$\pm$0.2\hspace{0.6cm} \\
A24 &  16$^{\mathrm h}$26$^{\mathrm m}$46\fs4 & -24$\degr$11$\arcmin$54$^{\prime\prime}$ &16$^{\mathrm h}$23$^{\mathrm m}$45\fs2 & -24$\degr$05$\arcmin$12$^{\prime\prime}$ & 3 &   57.1 &  3.0$\pm$0.4\hspace{0.6cm} \\
A25 &  16$^{\mathrm h}$26$^{\mathrm m}$49\fs2 & -24$\degr$20$\arcmin$02$^{\prime\prime}$ &16$^{\mathrm h}$23$^{\mathrm m}$47\fs9 & -24$\degr$13$\arcmin$19$^{\prime\prime}$ & 3 &   18.0 &  1.1$\pm$0.2\hspace{0.6cm} \\
A26 &  16$^{\mathrm h}$27$^{\mathrm m}$09\fs9 & -24$\degr$19$\arcmin$15$^{\prime\prime}$ &16$^{\mathrm h}$24$^{\mathrm m}$08\fs6 & -24$\degr$12$\arcmin$34$^{\prime\prime}$ & 6 &    9.7 &  1.2$\pm$0.3\hspace{0.6cm} \\
\hline
\hline
\vspace{-0.25cm}\\
F1~ &  16$^{\mathrm h}$26$^{\mathrm m}$23\fs5 & -24$\degr$43$\arcmin$10$^{\prime\prime}$ &  16$^{\mathrm h}$23$^{\mathrm m}$21\fs7 & -24$\degr$36$\arcmin$26$^{\prime\prime}$ & 3 &   54.8 & 2.2 $\pm$ 0.7\hspace{0.4cm} & $\leq$1.2\hspace{.82cm}\hspace{0.35cm} & 3.3 $\pm$ 0.4\hspace{0.4cm} &  2.0 $\pm$ 0.3\hspace{0.4cm} \\
F2~ &  16$^{\mathrm h}$26$^{\mathrm m}$44\fs0 & -24$\degr$43$\arcmin$14$^{\prime\prime}$ &  16$^{\mathrm h}$23$^{\mathrm m}$42\fs2 & -24$\degr$36$\arcmin$31$^{\prime\prime}$ & 2 &  126.9 & $\leq$1.4\hspace{.82cm}\hspace{0.35cm}  	& 1.8 $\pm$ 0.4\hspace{0.4cm} & 3.1 $\pm$ 0.3\hspace{0.4cm} & 2.1 $\pm$ 0.2\hspace{0.4cm} \\
F3~ &  16$^{\mathrm h}$26$^{\mathrm m}$46\fs8 & -24$\degr$44$\arcmin$28$^{\prime\prime}$ &  16$^{\mathrm h}$23$^{\mathrm m}$45\fs0 & -24$\degr$37$\arcmin$46$^{\prime\prime}$ & 2 &   25.5 & $\leq$0.8\hspace{.82cm}\hspace{0.35cm}  	& $\leq$0.7\hspace{.82cm}\hspace{0.35cm}  	&1.2 $\pm$ 0.2\hspace{0.4cm} & 0.7 $\pm$ 0.1\hspace{0.4cm} \\
F4~ &  16$^{\mathrm h}$26$^{\mathrm m}$47\fs8 & -24$\degr$43$\arcmin$05$^{\prime\prime}$ &  16$^{\mathrm h}$23$^{\mathrm m}$46\fs1 & -24$\degr$36$\arcmin$23$^{\prime\prime}$ & 3 &   7.1 &        	&         &         & 0.3 $\pm$ 0.1\hspace{0.4cm}  \\
F5~ &  16$^{\mathrm h}$26$^{\mathrm m}$50\fs8 & -24$\degr$36$\arcmin$43$^{\prime\prime}$ &  16$^{\mathrm h}$23$^{\mathrm m}$49\fs1 & -24$\degr$30$\arcmin$01$^{\prime\prime}$ & 3 &    6.8 & $\leq$1.0\hspace{.82cm}\hspace{0.35cm}  	& 0.5 $\pm$ 0.2\hspace{0.4cm} & $\leq$0.3\hspace{.82cm}\hspace{0.35cm} & $\leq$0.2\hspace{.82cm}\hspace{0.35cm} \\
F6~ &  16$^{\mathrm h}$26$^{\mathrm m}$56\fs9 & -24$\degr$38$\arcmin$42$^{\prime\prime}$ &  16$^{\mathrm h}$23$^{\mathrm m}$55\fs1 & -24$\degr$32$\arcmin$00$^{\prime\prime}$ & 2 &    9.1 & $\leq$0.5\hspace{.82cm}\hspace{0.35cm}  	& $\leq$0.4\hspace{.82cm}\hspace{0.35cm} 	& 0.4 $\pm$ 0.1\hspace{0.4cm} & 0.2 $\pm$ 0.1\hspace{0.4cm} \\
F7~ &  16$^{\mathrm h}$26$^{\mathrm m}$58\fs4 & -24$\degr$45$\arcmin$36$^{\prime\prime}$ &  16$^{\mathrm h}$23$^{\mathrm m}$56\fs5 & -24$\degr$38$\arcmin$55$^{\prime\prime}$ & 2 &   60.1 & 0.8 $\pm$ 0.3\hspace{0.4cm}  	& 2.3 $\pm$ 0.3\hspace{0.4cm} & 1.0 $\pm$ 0.2\hspace{0.4cm} & 1.3 $\pm$ 0.1\hspace{0.4cm} \\
F8~ &  16$^{\mathrm h}$27$^{\mathrm m}$04\fs3 & -24$\degr$43$\arcmin$00$^{\prime\prime}$ &  16$^{\mathrm h}$24$^{\mathrm m}$02\fs5 & -24$\degr$36$\arcmin$19$^{\prime\prime}$ & 2 &   11.4 & $\leq$0.9\hspace{.82cm}\hspace{0.35cm}  	& $\leq$0.7\hspace{.82cm}\hspace{0.35cm} 	& 0.4 $\pm$ 0.1\hspace{0.4cm} &  0.4 $\pm$ 0.1\hspace{0.4cm} \\
F9~ &  16$^{\mathrm h}$27$^{\mathrm m}$04\fs5 & -24$\degr$42$\arcmin$13$^{\prime\prime}$ &  16$^{\mathrm h}$24$^{\mathrm m}$02\fs7 & -24$\degr$35$\arcmin$32$^{\prime\prime}$ & 2 &   25.6 & 0.7 $\pm$ 0.3\hspace{0.4cm}  	& 0.4 $\pm$ 0.2\hspace{0.4cm} & 1.0 $\pm$ 0.2\hspace{0.4cm} & 0.8 $\pm$ 0.1\hspace{0.4cm} \\
F10 &  16$^{\mathrm h}$27$^{\mathrm m}$05\fs6 & -24$\degr$37$\arcmin$14$^{\prime\prime}$ &  16$^{\mathrm h}$24$^{\mathrm m}$04\fs0 & -24$\degr$30$\arcmin$33$^{\prime\prime}$ & 3 &   8.8 &  	&   &   & 0.2 $\pm$ 0.1\hspace{0.4cm}  \\
\hline
\vspace{-0.25cm}\\
F11 &  16$^{\mathrm h}$27$^{\mathrm m}$08\fs9 & -24$\degr$32$\arcmin$11$^{\prime\prime}$ &  16$^{\mathrm h}$24$^{\mathrm m}$07\fs3 & -24$\degr$25$\arcmin$29$^{\prime\prime}$ & 3 &    6.9 & $\leq$1.2\hspace{.82cm}\hspace{0.35cm}  	& 0.5 $\pm$ 0.2\hspace{0.4cm} & $\leq$0.3\hspace{.82cm}\hspace{0.35cm} & $\leq$0.2\hspace{.82cm}\hspace{0.35cm} \\
F12 &  16$^{\mathrm h}$27$^{\mathrm m}$09\fs7 & -24$\degr$49$\arcmin$02$^{\prime\prime}$ &  16$^{\mathrm h}$24$^{\mathrm m}$07\fs8 & -24$\degr$42$\arcmin$21$^{\prime\prime}$ & 4 &   8.9 &        	&         &         & 0.4 $\pm$ 0.1\hspace{0.4cm} \\
F13 &  16$^{\mathrm h}$27$^{\mathrm m}$10\fs1 & -24$\degr$36$\arcmin$50$^{\prime\prime}$ &  16$^{\mathrm h}$24$^{\mathrm m}$08\fs4 & -24$\degr$30$\arcmin$09$^{\prime\prime}$ & 2 &    8.9 & $\leq$0.6\hspace{.82cm}\hspace{0.35cm}  	& $\leq$0.2\hspace{.82cm}\hspace{0.35cm} 	& 0.3 $\pm$ 0.1\hspace{0.4cm} & $\leq$0.2\hspace{.82cm}\hspace{0.35cm} \\
F14 &  16$^{\mathrm h}$27$^{\mathrm m}$14\fs9 & -24$\degr$51$\arcmin$40$^{\prime\prime}$ &  16$^{\mathrm h}$24$^{\mathrm m}$12\fs9 & -24$\degr$44$\arcmin$59$^{\prime\prime}$ & 2 &  140.2 & 3.6 $\pm$ 0.7\hspace{0.4cm}   & 4.4 $\pm$ 0.5\hspace{0.4cm} & 4.3  $\pm$ 0.4\hspace{0.4cm} &  4.2 $\pm$ 0.3\hspace{0.4cm} \\
F15 &  16$^{\mathrm h}$27$^{\mathrm m}$15\fs5 & -24$\degr$30$\arcmin$49$^{\prime\prime}$ &  16$^{\mathrm h}$24$^{\mathrm m}$13\fs9 & -24$\degr$24$\arcmin$08$^{\prime\prime}$ & 6 &    7.1 & $\leq$0.9\hspace{.82cm}\hspace{0.35cm}   	& $\leq$0.4\hspace{.82cm}\hspace{0.35cm} 	& 1.0 $\pm$ 0.3\hspace{0.4cm} &  $\leq$0.3\hspace{.82cm}\hspace{0.35cm} \\
F16 &  16$^{\mathrm h}$27$^{\mathrm m}$15\fs8 & -24$\degr$38$\arcmin$42$^{\prime\prime}$ &  16$^{\mathrm h}$24$^{\mathrm m}$14\fs1 & -24$\degr$32$\arcmin$01$^{\prime\prime}$ & 2 &   15.1 & $\leq$0.5\hspace{.82cm}\hspace{0.35cm}  	& 0.7 $\pm$ 0.2\hspace{0.4cm} & 0.3 $\pm$ 0.1\hspace{0.4cm} &  0.5 $\pm$ 0.1\hspace{0.4cm} \\
F17 &  16$^{\mathrm h}$27$^{\mathrm m}$19\fs5 & -24$\degr$41$\arcmin$40$^{\prime\prime}$ &  16$^{\mathrm h}$24$^{\mathrm m}$17\fs6 & -24$\degr$35$\arcmin$00$^{\prime\prime}$ & 1 & 1648.5 & 16.6 $\pm$ 1.2\hspace{0.4cm}  & 13.8 $\pm$ 0.7\hspace{0.4cm} & 14.4 $\pm$ 0.6\hspace{0.4cm} &  13.2 $\pm$ 0.4\hspace{0.4cm} \\
F18 &  16$^{\mathrm h}$27$^{\mathrm m}$20\fs7 & -24$\degr$38$\arcmin$23$^{\prime\prime}$ &  16$^{\mathrm h}$24$^{\mathrm m}$19\fs0 & -24$\degr$31$\arcmin$43$^{\prime\prime}$ & 2 &    6.9 & $\leq$0.2\hspace{.82cm}\hspace{0.35cm}   	& $\leq$0.1\hspace{.82cm}\hspace{0.35cm} 	& 0.3 $\pm$ 0.1\hspace{0.4cm} & $\leq$0.1\hspace{.82cm}\hspace{0.35cm} \\
F19 &  16$^{\mathrm h}$27$^{\mathrm m}$22\fs9 & -24$\degr$48$\arcmin$07$^{\prime\prime}$ &  16$^{\mathrm h}$24$^{\mathrm m}$20\fs9 & -24$\degr$41$\arcmin$27$^{\prime\prime}$ & 2 &   12.6 & $\leq$0.5\hspace{.82cm}\hspace{0.35cm}  	& $\leq$0.3\hspace{.82cm}\hspace{0.35cm} 	& 0.5 $\pm$ 0.2\hspace{0.4cm} &  $\leq$0.3\hspace{.82cm}\hspace{0.35cm} \\
F20 &  16$^{\mathrm h}$27$^{\mathrm m}$24\fs8 & -24$\degr$41$\arcmin$24$^{\prime\prime}$ &  16$^{\mathrm h}$24$^{\mathrm m}$23\fs0 & -24$\degr$34$\arcmin$44$^{\prime\prime}$ & 2 &    7.2 & $\leq$0.3\hspace{.82cm}\hspace{0.35cm}  	& 0.3 $\pm$  0.1\hspace{0.4cm} & $\leq$0.2\hspace{.82cm}\hspace{0.35cm} & $\leq$0.2\hspace{.82cm}\hspace{0.35cm} \\
\hline
\vspace{-0.25cm}\\
F21 &  16$^{\mathrm h}$27$^{\mathrm m}$26\fs9 & -24$\degr$40$\arcmin$49$^{\prime\prime}$ &  16$^{\mathrm h}$24$^{\mathrm m}$25\fs0 & -24$\degr$34$\arcmin$09$^{\prime\prime}$ & 2 &   62.0 &  3.2 $\pm$ 0.6\hspace{0.4cm}  & $\leq$0.1\hspace{.82cm}\hspace{0.35cm} 	 & $\leq$0.2\hspace{.82cm}\hspace{0.35cm} &  0.7 $\pm$ 0.1\hspace{0.4cm} \\
F22 &  16$^{\mathrm h}$27$^{\mathrm m}$27\fs4 & -24$\degr$31$\arcmin$15$^{\prime\prime}$ &  16$^{\mathrm h}$24$^{\mathrm m}$25\fs8 & -24$\degr$24$\arcmin$35$^{\prime\prime}$ & 2 &  482.8 & $\leq$1.4\hspace{.82cm}\hspace{0.35cm}  	& 11.1 $\pm$ 0.7\hspace{0.4cm} & 0.6 $\pm$ 0.2\hspace{0.4cm} & 4.3 $\pm$ 0.3\hspace{0.4cm} \\
F23 &  16$^{\mathrm h}$27$^{\mathrm m}$28\fs1 & -24$\degr$48$\arcmin$04$^{\prime\prime}$ &  16$^{\mathrm h}$24$^{\mathrm m}$26\fs2 & -24$\degr$41$\arcmin$25$^{\prime\prime}$ & 3 &   7.1 &        	&         &         & 0.2 $\pm$ 0.1\hspace{0.4cm} \\
F24 &  16$^{\mathrm h}$27$^{\mathrm m}$28\fs6 & -24$\degr$54$\arcmin$31$^{\prime\prime}$ &  16$^{\mathrm h}$24$^{\mathrm m}$26\fs5 & -24$\degr$47$\arcmin$51$^{\prime\prime}$ & 5 &   12.8 & 2.4 $\pm$ 0.6\hspace{0.4cm}  	& $\leq$1.2\hspace{.82cm}\hspace{0.35cm}	& 1.4 $\pm$ 0.3\hspace{0.4cm} &  1.2 $\pm$ 0.2\hspace{0.4cm} \\
F25 &  16$^{\mathrm h}$27$^{\mathrm m}$30\fs1 & -24$\degr$31$\arcmin$55$^{\prime\prime}$ &  16$^{\mathrm h}$24$^{\mathrm m}$28\fs4 & -24$\degr$25$\arcmin$15$^{\prime\prime}$ & 3 &    7.3 & $\leq$0.6\hspace{.82cm}\hspace{0.35cm} 	& 0.5 $\pm$ 0.2\hspace{0.4cm} & $\leq$0.3\hspace{.82cm}\hspace{0.35cm} & $\leq$0.3\hspace{.82cm}\hspace{0.35cm} \\
F26 &  16$^{\mathrm h}$27$^{\mathrm m}$31\fs1 & -24$\degr$47$\arcmin$30$^{\prime\prime}$ &  16$^{\mathrm h}$24$^{\mathrm m}$29\fs1 & -24$\degr$40$\arcmin$50$^{\prime\prime}$ & 2 &   13.2 & 0.8 $\pm$ 0.3\hspace{0.4cm}  	& 0.6 $\pm$ 0.2\hspace{0.4cm} & $\leq$0.3\hspace{.82cm}\hspace{0.35cm} &  0.7 $\pm$ 0.1\hspace{0.4cm} \\
F27 &  16$^{\mathrm h}$27$^{\mathrm m}$33\fs2 & -24$\degr$41$\arcmin$14$^{\prime\prime}$ &  16$^{\mathrm h}$24$^{\mathrm m}$31\fs4 & -24$\degr$34$\arcmin$35$^{\prime\prime}$ & 2 &   48.6 & $\leq$0.7\hspace{.82cm}\hspace{0.35cm}  	& 0.9 $\pm$ 0.2\hspace{0.4cm} & 1.3 $\pm$ 0.2\hspace{0.4cm} &  1.0 $\pm$ 0.1\hspace{0.4cm} \\
F28 &  16$^{\mathrm h}$27$^{\mathrm m}$35\fs5 & -24$\degr$45$\arcmin$32$^{\prime\prime}$ &  16$^{\mathrm h}$24$^{\mathrm m}$33\fs7 & -24$\degr$38$\arcmin$53$^{\prime\prime}$ & 3 &   7.4 &        	&         &         & 0.2 $\pm$ 0.1\hspace{0.4cm} \\
F29 &  16$^{\mathrm h}$27$^{\mathrm m}$38\fs3 & -24$\degr$36$\arcmin$58$^{\prime\prime}$ &  16$^{\mathrm h}$24$^{\mathrm m}$36\fs6 & -24$\degr$30$\arcmin$19$^{\prime\prime}$ & 2 &  333.4 &  9.6 $\pm$ 0.9\hspace{0.4cm}  & 0.5 $\pm$ 0.2\hspace{0.4cm} & 0.5 $\pm$ 0.1\hspace{0.4cm} & 1.9 $\pm$ 0.2\hspace{0.4cm} \\
F30 &  16$^{\mathrm h}$27$^{\mathrm m}$39\fs5 & -24$\degr$39$\arcmin$15$^{\prime\prime}$ &  16$^{\mathrm h}$24$^{\mathrm m}$37\fs7 & -24$\degr$32$\arcmin$35$^{\prime\prime}$ & 1 &   61.5 & 1.5 $\pm$ 0.4\hspace{0.4cm}  	& 1.4 $\pm$ 0.3\hspace{0.4cm} & 1.1 $\pm$ 0.2\hspace{0.4cm} & 1.2 $\pm$ 0.1\hspace{0.4cm} \\
\hline
\vspace{-0.25cm}\\
F31 &  16$^{\mathrm h}$27$^{\mathrm m}$40\fs4 & -24$\degr$22$\arcmin$05$^{\prime\prime}$ &  16$^{\mathrm h}$24$^{\mathrm m}$39\fs0 & -24$\degr$15$\arcmin$26$^{\prime\prime}$ & 3 &   434.4 &  10.1 $\pm$ 1.1\hspace{0.4cm}  & 2.1 $\pm$ 0.5\hspace{0.4cm} 	& 12.3 $\pm$ 0.7\hspace{0.4cm} & 7.3 $\pm$ 0.4\hspace{0.4cm} \\
F32 &  16$^{\mathrm h}$27$^{\mathrm m}$51\fs9 & -24$\degr$46$\arcmin$31$^{\prime\prime}$ &  16$^{\mathrm h}$24$^{\mathrm m}$50\fs0 & -24$\degr$39$\arcmin$53$^{\prime\prime}$ & 2 &   7.7 &        	&         &         & 0.3 $\pm$ 0.1\hspace{0.4cm} \\
F33 &  16$^{\mathrm h}$27$^{\mathrm m}$52\fs1 & -24$\degr$40$\arcmin$49$^{\prime\prime}$ &  16$^{\mathrm h}$24$^{\mathrm m}$50\fs3 & -24$\degr$34$\arcmin$11$^{\prime\prime}$ & 1 &  199.4 & 2.9 $\pm$ 0.5\hspace{0.4cm}  	& 2.7 $\pm$ 0.3\hspace{0.4cm} & 2.9 $\pm$ 0.3\hspace{0.4cm} & 2.8 $\pm$ 0.2\hspace{0.4cm} \\
F34 &  16$^{\mathrm h}$27$^{\mathrm m}$59\fs3 & -24$\degr$48$\arcmin$19$^{\prime\prime}$ &  16$^{\mathrm h}$24$^{\mathrm m}$57\fs3 & -24$\degr$41$\arcmin$42$^{\prime\prime}$ & 5 &    7.0 & $\leq$1.2\hspace{.82cm}\hspace{0.35cm}  	& 0.8 $\pm$ 0.3\hspace{0.4cm} & $\leq$0.6\hspace{.82cm}\hspace{0.35cm} & $\leq$0.5\hspace{.82cm}\hspace{0.35cm} \\
F35 &  16$^{\mathrm h}$28$^{\mathrm m}$00\fs1 & -24$\degr$53$\arcmin$44$^{\prime\prime}$ &  16$^{\mathrm h}$24$^{\mathrm m}$58\fs0 & -24$\degr$47$\arcmin$07$^{\prime\prime}$ & 3 &   59.4 & $\leq$2.8\hspace{.82cm}\hspace{0.35cm} 	& $\leq$1.9\hspace{.82cm}\hspace{0.35cm} 	& 3.7 $\pm$ 0.5\hspace{0.4cm} & 2.5 $\pm$ 0.3\hspace{0.4cm} \\
F36 &  16$^{\mathrm h}$28$^{\mathrm m}$12\fs3 & -24$\degr$50$\arcmin$47$^{\prime\prime}$ &  16$^{\mathrm h}$25$^{\mathrm m}$10\fs2 & -24$\degr$44$\arcmin$10$^{\prime\prime}$ & 4 &   31.9 &  $\leq$2.2\hspace{.82cm}\hspace{0.35cm}  	& $\leq$1.6\hspace{.82cm}\hspace{0.35cm} 	& 2.4 $\pm$ 0.4\hspace{0.4cm} & 1.7 $\pm$ 0.2\hspace{0.4cm} \\
F37 &  16$^{\mathrm h}$28$^{\mathrm m}$25\fs3 & -24$\degr$45$\arcmin$00$^{\prime\prime}$ &  16$^{\mathrm h}$25$^{\mathrm m}$23\fs3 & -24$\degr$38$\arcmin$24$^{\prime\prime}$ & 3 &   47.4 & 2.5 $\pm$ 0.6\hspace{0.4cm} 	& 3.5 $\pm$ 0.5\hspace{0.4cm} & 1.8 $\pm$ 0.3\hspace{0.4cm} & 2.6 $\pm$ 0.3\hspace{0.4cm} \\
\hline
\hline
\end{tabular}
\label{tab:sources}
\normalsize
\vspace{0.2cm}\\
{\noindent {\bf Notes:}  
$\pm$ gives the $1 \sigma_{total}$ error box; $\cal L$ is the likelihood of existence, we give the maximum
value for the observation set.}
\end{table*}

In order to allow easier comparisons with previous work, 
\xr source positions are
listed in both J2000 and B1950 equinoxes, with their $1 \sigma_{total}$ error
box, in Cols.~2--6.  The likelihood of existence $\cal L$ is in Col.~7.
Count rates are indicated in Col.~8--11.  For the core~F field, the
indicated positions ($\alpha, \delta$) and $\cal L$ values correspond to
observation (\#1, \#2, or \#3) where $\cal L$ and the position accuracy are the
best, i.e.  when the count rate is highest.  When an \xr source is detected in
one observation above the detection threshold, and not detected in other
observations, we have estimated the corresponding count rate upper limits (3.25$\sigma$), 
using the EXSAS command COMPUTE/UPPER\_LIMITS.  We have noted
that the detection efficiency degrades with increasing angle to the axis, in the
same way as the point spread function (this is discussed in $\S$\ref{X_versus_IR}).

\section{Optical/IR counterparts of the HRI \xr sources}
\label{appendix:finding_map}

We searched stellar counterparts for the 63 \ROSAT \HRI \xr sources on the ESO/SERC second 
digitized sky survey (DSS2).
Fig.~\ref{finding_map} gives the finding charts with BKLT IR sources for each of the 63 \ROSAT \HRI \xr sources.

Table~\ref{tab:ident} gives identification lists for the two fields, 
and cross-identification with other surveys.
Col.~1 is the ROXR numbering from detection (Table~\ref{tab:sources}).
Cols.~2--4 are respectively cross-identification lists with the \xr sources
of CMFA (ROXR1), Casanova (\cite{casanova94}; ROXR2), and Kamata et~al. (\cite{kamata97}).  
Dots mean ``\xr source undetected'', and dash ``out of observation field''.  
Col.~5 gives the first name attributed to this counterpart.

In the core~A field, we find 26 \xr sources, of which only one (ROXRA10)
remains without optical or IR counterpart.  Of the 25 identified \xr sources, 22
were seen with the \ROSAT {\it PSPC}, and 4 are new detections (ROXRA3, 10, 16,
22).  In the core~F field, we find 37 \xr sources, including 7 without optical
or IR counterpart.  Of the 30 identified \xr sources, 18 were seen with the
\ROSAT {\it PSPC}, and 12 are new detections (ROXRF3, 8, 12, 15, 18, 19, 24,
26, 28, 32, 35, 36).
Altogether, 63 \xr sources are detected, and 55 are identified.
Of the 55 identified \xr sources 40 are \PSPC sources.

For sources with a low statistical significance ($6.8 \le {\cal L} \le 9.1$, 
or $10^{-4} \le  P_0 \le 1.1\times 10^{-3}$; 3.25--3.9\,$\sigma$ for Gaussian statistics) 
we find \xr sources with and without optical or IR counterparts.  
The \xr sources without counterparts are
always weak sources and may be spurious detections (locally high background),
and this may therefore also be the case for weak \xr sources with counterparts
in case of chance spatial coincidence.  For instance in the Core~A field
(respectively Core~F) there are 875 (resp.  1173) BKLT sources; this sample is
dominated by background sources without detectable \xr emission.  To estimate
the number of chance coincidences, we have placed in each field 10$^5$ random X-ray
source positions, and searched for each whether there is a
BKLT source in a circle of 10$\arcsec$ radius:  we have then an estimate of the
probability to find a BKLT counterpart by chance within 10$\arcsec$ from a spurious
\xr detection.  This probability is 0.044 (resp.  0.049) for Core~A (resp.  Core
F), or approximately 1/20 for both; in other words one (resp.  two) spurious source identification
are expected for the Core~A (resp.  Core~F) field.  As we have for Core~A (resp.
Core~F) two (resp.  15) \xr sources with ${\cal L} \le 9.1$ out of 26 (resp.  out
of 37), this implies that one weak X-ray source in Core~A (resp.  13 in Core~B) is
real, which is consistent with the number of identifications.  
We are therefore confident that the identifications of weak X-ray sources with 
stellar counterparts are correct.


\begin{figure*}
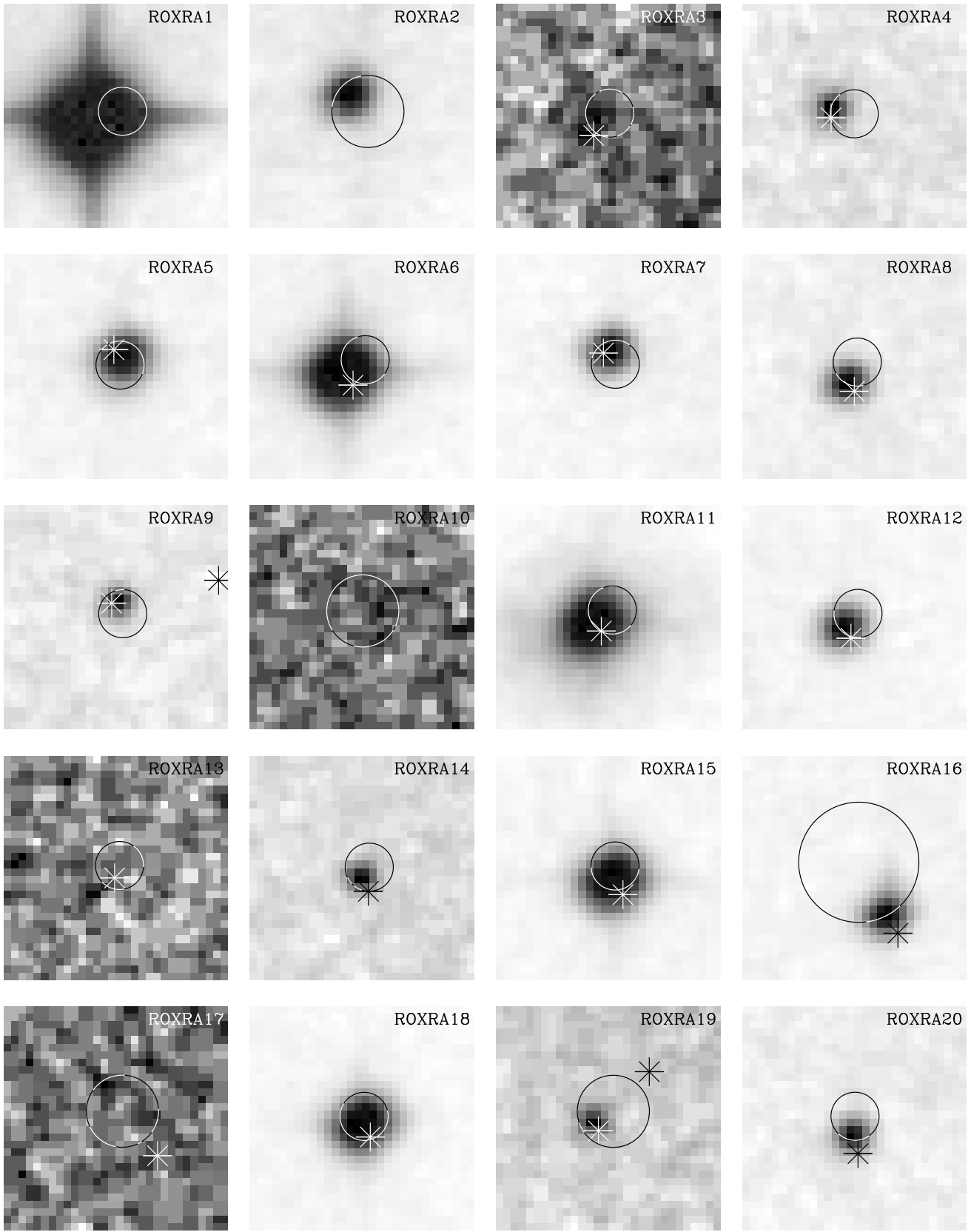
 
\begin{center}
\begin{tabular}{ccccc}
\resizebox{4cm}{!}{\includegraphics{9083_fb1.a1}} & \resizebox{4cm}{!}{\includegraphics{9083_fb1.a2}} & \resizebox{4cm}{!}{\includegraphics{9083_fb1.a3}} &  \resizebox{4cm}{!}{\includegraphics{9083_fb1.a4}} \\
\\
\resizebox{4cm}{!}{\includegraphics{9083_fb1.a5}} & \resizebox{4cm}{!}{\includegraphics{9083_fb1.a6}} & \resizebox{4cm}{!}{\includegraphics{9083_fb1.a7}} &  \resizebox{4cm}{!}{\includegraphics{9083_fb1.a8}} \\
\\
\resizebox{4cm}{!}{\includegraphics{9083_fb1.a9}} & \resizebox{4cm}{!}{\includegraphics{9083_fb1.a10}} & \resizebox{4cm}{!}{\includegraphics{9083_fb1.a11}} &  \resizebox{4cm}{!}{\includegraphics{9083_fb1.a12}} \\
\\
\resizebox{4cm}{!}{\includegraphics{9083_fb1.a13}} & \resizebox{4cm}{!}{\includegraphics{9083_fb1.a14}} & \resizebox{4cm}{!}{\includegraphics{9083_fb1.a15}} &  \resizebox{4cm}{!}{\includegraphics{9083_fb1.a16}} \\
\\
\resizebox{4cm}{!}{\includegraphics{9083_fb1.a17}} & \resizebox{4cm}{!}{\includegraphics{9083_fb1.a18}} & \resizebox{4cm}{!}{\includegraphics{9083_fb1.a19}} &  \resizebox{4cm}{!}{\includegraphics{9083_fb1.a20}}
\end{tabular}
\end{center}
\caption{Optical finding charts of the 63 \ROSAT \HRI \xr sources 
of Tables~\ref{tab:sources} and \ref{tab:ident}.
Each map is a $30\arcsec \times 30\arcsec$ extracted from the ESO/SERC sky survey 
red Schmidt plate using the second Digitized Sky Survey (one pixel=1$\arcsec$); 
North is at the top, East at the left.
Circles show the \ROSAT \HRI 90$\%$ confidence error boxes 
(i.e. one sigma error box from Table~\ref{tab:sources} multiplied by 1.6).
Asterisks show the BKLT infrared sources 90$\%$ confidence error boxes ($\sim 1.9\arcsec$).}
\label{finding_map}
\end{figure*}

\addtocounter{figure}{-1}
\begin{figure*} 
\begin{center}
\begin{tabular}{ccccc}
\resizebox{4cm}{!}{\includegraphics{9083_fb1.a21}} & \resizebox{4cm}{!}{\includegraphics{9083_fb1.a22}} & \resizebox{4cm}{!}{\includegraphics{9083_fb1.a23}} &  \resizebox{4cm}{!}{\includegraphics{9083_fb1.a24}} \\
\\
\resizebox{4cm}{!}{\includegraphics{9083_fb1.a25}} & \resizebox{4cm}{!}{\includegraphics{9083_fb1.a26}} & \resizebox{4cm}{!}{\includegraphics{9083_fb1.f1}} &  \resizebox{4cm}{!}{\includegraphics{9083_fb1.f2}} \\
\\
\resizebox{4cm}{!}{\includegraphics{9083_fb1.f3}} & \resizebox{4cm}{!}{\includegraphics{9083_fb1.f4}} & \resizebox{4cm}{!}{\includegraphics{9083_fb1.f5}} &  \resizebox{4cm}{!}{\includegraphics{9083_fb1.f6}} \\
\\
\resizebox{4cm}{!}{\includegraphics{9083_fb1.f7}} & \resizebox{4cm}{!}{\includegraphics{9083_fb1.f8}} & \resizebox{4cm}{!}{\includegraphics{9083_fb1.f9}} &  \resizebox{4cm}{!}{\includegraphics{9083_fb1.f10}} \\
\\
\resizebox{4cm}{!}{\includegraphics{9083_fb1.f11}} & \resizebox{4cm}{!}{\includegraphics{9083_fb1.f12}} & \resizebox{4cm}{!}{\includegraphics{9083_fb1.f13}} &  \resizebox{4cm}{!}{\includegraphics{9083_fb1.f14}}
\end{tabular}
\end{center}
\caption{\it continued.}
\end{figure*}

\addtocounter{figure}{-1}
\begin{figure*}
\begin{tabular}{ccccc}
\resizebox{4cm}{!}{\includegraphics{9083_fb1.f15}} & \resizebox{4cm}{!}{\includegraphics{9083_fb1.f16}} & \resizebox{4cm}{!}{\includegraphics{9083_fb1.f17}} &  \resizebox{4cm}{!}{\includegraphics{9083_fb1.f18}} \\
\\
\resizebox{4cm}{!}{\includegraphics{9083_fb1.f19}} & \resizebox{4cm}{!}{\includegraphics{9083_fb1.f20}} & \resizebox{4cm}{!}{\includegraphics{9083_fb1.f21}} &  \resizebox{4cm}{!}{\includegraphics{9083_fb1.f22}} \\
\\
\resizebox{4cm}{!}{\includegraphics{9083_fb1.f23}} & \resizebox{4cm}{!}{\includegraphics{9083_fb1.f24}} & \resizebox{4cm}{!}{\includegraphics{9083_fb1.f25}} &  \resizebox{4cm}{!}{\includegraphics{9083_fb1.f26}} \\
\\
\resizebox{4cm}{!}{\includegraphics{9083_fb1.f27}} & \resizebox{4cm}{!}{\includegraphics{9083_fb1.f28}} & \resizebox{4cm}{!}{\includegraphics{9083_fb1.f29}} &  \resizebox{4cm}{!}{\includegraphics{9083_fb1.f30}} \\
\\
\resizebox{4cm}{!}{\includegraphics{9083_fb1.f31}} & \resizebox{4cm}{!}{\includegraphics{9083_fb1.f32}} & \resizebox{4cm}{!}{\includegraphics{9083_fb1.f33}} &  \resizebox{4cm}{!}{\includegraphics{9083_fb1.f34}}
\end{tabular}
\caption{\it continued.}
\end{figure*}

\addtocounter{figure}{-1}
\begin{figure*} 
\begin{tabular}{cccc}
\resizebox{4cm}{!}{\includegraphics{9083_fb1.f35}} & \resizebox{4cm}{!}{\includegraphics{9083_fb1.f36}} & \resizebox{4cm}{!}{\includegraphics{9083_fb1.f37}}
\end{tabular}
\caption{\it continued.} 
\end{figure*}

\section{Comparison between \HRI and \PSPC observations}
\label{Comparison_HRI_PSPC}

Within the boundaries of our observation fields (core~A and core~F), there are
61 \xr sources detected previously with the \ROSAT \PSPC (53 from CMFA, and 8
from Casanova \cite{casanova94}).  However, 21 \xr sources are not detected with the \ROSAT
{\it HRI}.  This difference could be explained by lower observational sensitivity
and/or source variability.  To elucidate this point, we must estimate \HRI count
rates for \PSPC sources, and compare them with the adopted \HRI detection threshold 
(3.25\,$\sigma$).

We first estimate the conversion factor between the \PSPC count rate in the
energy range of CMFA (1.0--2.4\,keV) and the \HRI count rate in the whole energy
range (0.1--2.4\,keV).  
We did not select \xr sources with ambiguous \PSPC detection 
(10 sources with notes in Cols.~2--3 of Table~\ref{tab:ident}).
For core~F observation, the lowest detection count rate was taken to minimize
variability effects (four sources with upper limits are not selected): 
we kept only 26 \xr sources.   
Since many of these sources are
variable (as shown by the core~F observations), a conversion
factor estimator insensitive to extreme values of the sample is needed.  
This is why we take the median of the {\it PSPC}/\HRI count rate ratio, instead of the mean.  
We find  \PSPC$_{\mathrm{1-2.4\,keV}}$ count rate = 2.4 $\times$ 
\HRI$_{\mathrm{0.1-2.4\,keV}}$ count rate.

\addtocounter{section}{-1}
\begin{table*} 
\caption{{\it HRI} X-ray sources counterpart identifications in the $\rho$ Oph cloud cores A \& F.}
\scriptsize
\begin{tabular}{crrclcrrrcrrr}
\hline
\hline
\vspace{-0.1cm}\\
\multicolumn{3}{c}{ROXR}  &  & &  & &  & &  \multicolumn{4}{c}{$\log $L$_{\mathrm X}$~(d=140\,pc)}\\
\multicolumn{3}{c}{\hrulefill} & \multicolumn{1}{c}{ASCA} & \multicolumn{1}{c}{NAME} & ISO & \multicolumn{1}{c}{IR} &\multicolumn{1}{c}{A$_{\mathrm{V}}$} & \multicolumn{1}{c}{L$_{\star}$} & \multicolumn{4}{c}{\hrulefill} \\
 & \multicolumn{1}{c}{1} & \multicolumn{1}{c}{2} & &  & & \multicolumn{1}{c}{Class} & \multicolumn{1}{c}{[mag]}
 & \multicolumn{1}{c}{[L$_{\odot}$]} & \multicolumn{1}{c}{\#1} & \multicolumn{1}{c}{\#2} & \multicolumn{1}{c}{\#3} & \multicolumn{1}{c}{\#1+2+3}\\
(1) & (2) & (3) & \multicolumn{1}{c}{(4)} & \multicolumn{1}{c}{(5)} & \multicolumn{1}{c}{(6)} & \multicolumn{1}{c}{(7)} & \multicolumn{1}{c}{(8)} & \multicolumn{1}{c}{(9)} & \multicolumn{1}{c}{(10)} & \multicolumn{1}{c}{(11)} & \multicolumn{1}{c}{(12)} & \multicolumn{1}{c}{(13)} \\

\hline
\vspace{-0.25cm}\\
A1~ &  2~    			&   ---~  					& ---~		& \object{SR2}=\object{SAO184375}		&  --- 		&   III$^{\star}$\hspace{-0.055cm} 	&  0$^{\star}$\hspace{0.18cm} 	&	1.2$^{\star}$\hspace{-0.14cm}				& 30.4    \\
A2~ &  3~    			&   ---~  					& \dotfill~		&  \object{SR22}         	&  --- 		&   II$^{\star}$\hspace{-0.055cm} 	&  1.5$^{\star}$\hspace{-0.04cm}	&	0.4$^{\star}$\hspace{-0.14cm} & 29.4    \\
A3~ & \dotfill~  		&   ---~  					& \dotfill~ 	&  \object{B162538-242238} 		&   {\it red}  	&   nII~	 			&  10.9~				& 	0.2    			& 30.3    \\
A4~ &  5$^{\star}$\hspace{-0.04cm} 	&   ---~    					& \dotfill~	&  \object{IRS3}         	&   {\it red}	&   nII~	 			&  4.6~					&	0.3  				& 29.8    \\
A5~ &  8~    			&   ---~  					& 1		&  \object{ROXs4}=\object{IRS10}         		&  {\it blue}	&   nIII~	 			&  5.4~					&	1.7    			& 31.4    \\
A6~ &  9~    			&   ---~  					& \dotfill~	&  \object{SR4}          		&   {\it red}  	&   II~   	 	&  1.9~ 				&	1.5    			& 29.9   \\
A7~ & 10~    			&   ---~  					& 2		&  \object{GSS20}         		&  {\it blue}	&   III~    	 		&  4.7~					&	1.1  				& 30.5    \\
A8~ & 11~		 	&   ---~ 					& \dotfill~	&  \object{Chini8}$^{\star}$=\object{SKS3}       	&  \dotfill~	&  nIII$^{\star}$\hspace{-0.055cm}					&  0.5~				&  0.2				&  29.1 \\
A9~ & 12$^{\star}$\hspace{-0.04cm} 	&   ---~  				& \dotfill~	&  \object{B162601-242945}		&  {\it blue}  	&  nIII~	 			&  8.1~					&	0.5    			& 30.2   \\
A10 & \dotfill~  		&   ---~  					& \dotfill~	&  		              	&  \dotfill  	& 	 		&  				&			&  ?\\
\hline
\vspace{-0.25cm}\\
A11 & 13~    			&   ---~  					& 3		&  \object{DoAr21}=\object{ROXs8}              	&  {\it blue}  	&   III~	    	 &  6.0~					&	14.7   			& 31.6    \\
A12 & 14~		 	&   ---~  					& \dotfill~	&  \object{VSSG19}              		&  {\it blue}	&   nIII~	 			&  3.9~  				&	0.5    			& 29.9   \\
A13 & 15$^{\star}$\hspace{-0.04cm}	&   ---~  				& \dotfill~	&  \object{ROXC1}=\object{B162607-242742}	&  {\it blue}  	&  nIII~	 			&  20.6~				&	0.7    			& 31.3  \\
A14 & 18~    			&   ---~  					& \dotfill~	&  \object{GSS29}         		&   {\it red}	&   II~   	 			&  9.4~					&	1.4    			& 31.0    \\
A15 & 17~    			&   ---~  					& \dotfill~	&  \object{DoAr24}=\object{ROXs10A}    		&   {\it red} 	&   II~    	 			&  1.8~					&	0.7    			& 30.0    \\
A16 & ---~ 			& .\hspace{0.05cm}.$^{\star}$\hspace{-0.03cm} 	&  ---~		&  \object{WSB28}   		&  {\it blue} 	&  nIII~ 				&  4.3~					&	0.4				& 30.1\\
A17 & 22$^{\star}$\hspace{-0.04cm}   &   ---~  				& \dotfill~	&  \object{GSS30-IRS2} 			&  {\it blue}	&   III~    	 		& 19.5~					&	0.9    			& 30.9    \\
A18 & 24~    			&   ---~  					& 4		&  \object{GSS31}     			&   {\it red}  	&   II~    	 			&  6.1~					&	5.9    		& 30.4    \\
A19 & 25~    			&   ---~  					& C3		&  \object{S2}    			&   {\it red}	&   II~    	 			& 11.8~					&	3.7    			& 30.4    \\
A20 & 26~    			&   ---~  					& \dotfill~	&  \object{El24}      			&   {\it red}	&   II~    	 			& 10.0~					&	4.5    			& 30.5    \\
\hline
\vspace{-0.25cm}\\
A21 & 28~    			&   ---~  					& 5		&  \object{S1}=\object{ROXs14}        	&  {\it blue}  	&   	III$^{\star}$\hspace{-0.055cm}	    	&  12$^{\star}$\hspace{-0.04cm}					&   1100$^{\star}$\hspace{-0.04cm}	   		&  30.8   \\
A22 & \dotfill~  		&   ---~  					& \dotfill~	&  \object{B162636-241554} 		&   {\it red}  	&   nII~	 			&  7.6~					&	0.3     			& 30.2    \\
A23 & 29~    			&   ---~  					& C4		&  \object{GSS37}   			&   {\it red}	&   II~   	 			&  8.5~					&	1.6    			& 30.3    \\
A24 & ---~   			&   17~   					& \dotfill~	&  \object{VSS27}=\object{ROXs16} 		&   {\it red}	&   II~   	 			&  5.6~					&	1.9        			& 30.7    \\
A25 & 32~   			&   ---~  					& \dotfill~	&  \object{VSSG3}          		&  {\it blue}	&   III~    	 		& 15.7~					&	2.8    			& 31.2    \\
A26 & 38~   			&   ---~  					& \dotfill~	&  \object{SR21}$^{\star}$ 		&   {\it red} 	&   II~    	 			&  3.5~					&	4.0    			& 30.0    \\
\hline
\hline
\vspace{-0.25cm}\\
F1~ & 23~			&   ---~  					& \dotfill~	&  \object{DoAr25}$^{\star}$   		&  {\it red} 		&  II~		 			& 0.7~					&  0.8   	&  29.7  	& $\leq$29.4	& 29.8 		&	 29.6\hspace{0.45cm} \\
F2~ & 31~			&   ---~  					& \dotfill~	&  \object{GY112}    			& {\it blue} 		&nIII~		 			& 3.6~					&  0.4   	&  $\leq$30.0~~ & 30.2	     	& 30.4 		& 30.2\hspace{0.45cm} \\
F3~ & \dotfill~  		&   ---~  					& \dotfill~	&  \object{GY122}       			& {\it blue} 	& nIII~		 			& 2.7~					&  0.08   	&  $\leq$29.7~~ & $\leq$29.6 	& 29.8 		& 29.6\hspace{0.45cm} \\
F4~ & \dotfill~			&   ---~  					& \dotfill~	&  	   			&  \dotfill 	&  			& 				&  		& $\leq$	&  $\leq$    	& $\leq$	&  ?\hspace{0.45cm} \\
F5~ & \dotfill~  		&   ---~  					& \dotfill~	&  				&  \dotfill 	& 	 			& 				&  	&  ?	&  ?	&  ?	& ?\hspace{0.45cm} \\
F6~ & \dotfill~  		&   ---~  					& \dotfill~	&  	 			&  \dotfill 	& 	 			& 				&  	&  ?	&  ?	&  ?	& ?\hspace{0.45cm} \\
F7~ & 35$^{\star}$\hspace{-0.06cm}			&   ---~  		& \dotfill~	&  \object{SR24S}$^{\star}$    			&  {\it red} 	&  II~  	 			& 5.9~  				&  2.2  	&  30.1   	&  30.6		& 30.2 		& 30.3\hspace{0.45cm} \\
F8~ & \dotfill~  		&   ---~  					& \dotfill~	&  \object{GY193} 			& {\it blue} 	& nIII~		 			& 7.4~					&  0.3   	&  $\leq$30.3~~ & $\leq$30.2	& 30.0 		& 30.0\hspace{0.45cm} \\
F9~ & 36$^{\star}$\hspace{-0.06cm}     &   ---~  				& \dotfill~	&  \object{GY194} 		& {\it blue} 	& nIII~		 			& 9.1~					&  0.4   	&  30.4  	& 30.2		& 30.6 		& 30.5\hspace{0.45cm} \\
F10 & \dotfill~			&   ---~  					& \dotfill~	& 	   			&  \dotfill 	&	 			& 				&   	&  $\leq$	& $\leq$	&   $\leq$   	& ?\hspace{0.45cm} \\
\hline
\vspace{-0.25cm}\\
F11 & \dotfill~  		&   ---~  					& \dotfill~	&  	  			&  \dotfill 	& 	 			& 				&  	&  ?	& ?	&  ?	&  ?\hspace{0.45cm} \\
F12 & \dotfill~  		&   ---~  					& \dotfill~	&  {\it anonymous star}$^{\star}$ 	& ---	 	& II-III?\hspace{0cm}		& 				&  	&  $\leq$	& $\leq$	&  $\leq$    		& ?\hspace{0.45cm} \\
F13 & \dotfill~  		&   ---~  					& \dotfill~	&  	 			&  \dotfill 	& 	 			& 				&  	&  ?	& ?	&  ?	&  ?\hspace{0.45cm} \\
F14 & ---~ 			&   22$^{\star}$\hspace{-.06cm}   		& ---~		&  \object{ROXs20B} 		& --- 		& III~   	 			& 2.7~ 					&  0.3   	& 30.3   	&  30.4		& 30.4 		& 30.4\hspace{0.45cm} \\
F15 & \dotfill~  		&   ---~  					& \dotfill~	&  \object{GY238}			&  {\it red} 	& nII~		 		& 45.7~				&  0.3   	&  $\leq$32.9~~ & $\leq$32.6	& 33.0 		& $\leq$32.5\hspace{0.45cm} \\
F16 & 39~  			&   ---~  					& \dotfill~	&  \object{WL20}    			&  {\it red} 	& II~    	 			& 16.5~  				&  0.9   	&  $\leq$30.9~~ & 31.1		& 30.7 		& 30.9\hspace{0.45cm} \\
F17 & 41~     			&   ---~  					& 9A		&  \object{SR12AB}=\object{ROXs21} 			& {\it blue} 	& III~   	 			& 1.2~   				&  0.8   	&  30.7  	& 30.6		& 30.6 		& 30.6\hspace{0.45cm} \\
F18 & \dotfill~  		&   ---~  					& \dotfill~	&  \object{B162720-243820}		&  \dotfill 	& II-III?\hspace{-0.04cm}		& 				&  	&  ?	& ?	&  ?	& ?\hspace{0.45cm} \\
F19 & \dotfill~  		&   ---~  					& \dotfill~		&  \object{WSB49} 		&  \dotfill$^{\star}$		& II$^{\star}$\hspace{-0.055cm}	& 1.9~					&  0.2   	&  $\leq$29.3~~	& $\leq$29.1	& 29.3 		& $\leq$29.1\hspace{0.45cm} \\
F20 & \dotfill~  		&   ---~  					& \dotfill~	&  				&  \dotfill 	& 	 			& 				&  	&  ?	& ?	&  ?	& ?\hspace{0.45cm} \\
\hline
\vspace{-0.25cm}\\
F21 & 43$^{\star}$\hspace{-0.04cm} 	&   ---~  				& \dotfill~	&  \object{YLW15}=\object{IRS43}   	& {\it red} 	& I~   		 			& 30$^{\star}$\hspace{0.18cm}  		&10$^{\star}$\hspace{0.24cm} &  32.6 	& $\leq$31.1	&  $\leq$31.4 	& 31.9\hspace{0.45cm} \\
F22 & 44~  			&   ---~  					& \dotfill~	&  \object{VSSG25}    			& {\it red} 	& II~    	 			& 9.8~  				&  0.4   	&  $\leq$30.8~~	& 31.7		& 30.4 		& 31.3\hspace{0.45cm} \\
F23 & 46$^{\star}$\hspace{-0.06cm} 	&   ---~  				& \dotfill~	&  \object{B162728-244803}$^{\star}$   	& --- 		& II-III?\hspace{-0.04cm}		& 				&  	&  $\leq$	& $\leq$	&  $\leq$	& ?\hspace{0.45cm} \\
F24 & ---~       		&\dotfill~					& ---~		& \object{U0600-11613195}$^{\star}$ & --- 		& II-III?\hspace{-0.04cm}		& 				&  	&  ?	& ?	&  ?	& ?\hspace{0.45cm} \\
F25 & 48$^{\star}$\hspace{-0.02cm}     &   ---~  				& \dotfill~	&  \object{BBRCG50} 		&  \dotfill 	& II-III?\hspace{-0.04cm}		& 				&  	&  $\leq$	& $\leq$	& $\leq$	& ?\hspace{0.45cm} \\
F26 & \dotfill~  		&   ---~  					& \dotfill~	&   \object{B162730-244726} 		& {\it blue} 	& nIII~ 				& 10.3~					&  0.6   	& 30.6 		& 30.5		&  $\leq$30.2 	& 30.5\hspace{0.45cm} \\
F27 & 49~  			&   ---~  					& \dotfill~	&  \object{GY292}    			& {\it red} 	& II~    	 			& 10.8~					&  1.6   	& $\leq$30.6~~	& 30.7		& 30.8 		& 30.7\hspace{0.45cm} \\
F28 & \dotfill~			&   ---~  					& \dotfill~	&  \object{GY296}$^{\star}$ 		& {\it blue}	& nIII~	 			& 5.1~						&  0.08		&  	$\leq$	& $\leq$		&    $\leq$  		&  29.4\hspace{0.45cm} \\
F29 & 50~ 			&   ---~  					& \dotfill~	&  \object{IRS49}   			& {\it red} 	& II~  		 			& 10.1~ 				&  1.1  	& 31.6		& 30.4		& 30.4 		& 30.9\hspace{0.45cm} \\	
F30 & 51~			&   ---~  					& \dotfill~	&  \object{GY314}   			& {\it red} 	& II~    	 			& 6.4~   				&  0.8   	& 30.4 		& 30.4		& 30.3 		& 30.4\hspace{0.45cm} \\
\hline
\vspace{-0.25cm}\\
F31 & 52~     			&   ---~  					& 11		&  \object{SR9}=\object{ROXs29}    		& {\it red} 	& II~    	 			& 0~~~  				&  1.6   	& 29.8		& 29.2		& 29.9 		& 29.7\hspace{0.45cm} \\
F32 & \dotfill~			&   ---~  					& ---~		&  \object{GY377} 			&  {\it blue} 	& nIII~			& 	16.0~					&  0.4		&  	$\leq$	& $\leq$		&     $\leq$ 		&  30.6\hspace{0.45cm} \\
F33 & 54~     			&   ---~  					& \dotfill~	&  \object{ROXs31}=\object{IRS55}			&  {\it blue}	& III~   				& 6.0~  				&  1.8   	& 30.7		& 30.7		& 30.7 		& 30.7\hspace{0.45cm} \\
F34 & 55~  			&   ---~  					& ---~		&  \object{WSB58}    			& --- 		& II-III?\hspace{-0.04cm}		& 				&  	&  ?	& ?	&  ?	& ?\hspace{0.45cm} \\	
F35 & ---~       		&\dotfill~					& ---~		&  \object{B162800-245340}		& --- 		& II-III?\hspace{-0.04cm}		& 				&  	&  ?	& ?	&  ?	& ?\hspace{0.45cm} \\
F36 & ---~       		&\dotfill~					& ---~		&  \object{B162812-245043}		& --- 		& II-III?\hspace{-0.04cm}		& 				&  	&  ?	& ?	&  ?	& ?\hspace{0.45cm} \\
F37 & ---~       		&   34~   					& ---~		&  \object{HD148352}		& {\it blue} 	& III$^{\star}$\hspace{-0.055cm}						& 0~~~					&  	   	& 29.9		& 30.0		& 29.8 		& 29.9\hspace{0.45cm} \\
\hline
\hline
\end{tabular}
\normalsize
\vspace{0.2cm}\\
{\noindent{\bf Comments to Table~\ref{tab:ident}:}}\\
$\star$ = see notes below.
ROXRA or ROXRF = X-ray source number (this article). 
ROXR1 = CMFA X-ray source number. 
ROXR2 = Casanova (\cite{casanova94}) X-ray source number. 
ASCA = Kamata et~al. (\cite{kamata97}) X-ray source number. 
Dash = out of observation field. 
Dots = unobserved source. 
{\it red} = {\it ISOCAM} source with IR excess. 
{\it blue} = {\it ISOCAM} source without IR excess. 
nII = new class II. 
nIII = new class III. 
? = X-ray detected source for which intrinsic X-ray luminosity cannot be determined.
$\leq$ = X-ray undetected source. 
II-III? = class II or class III candidate (see Appendix B).
\label{tab:ident}
\end{table*}

\begin{table*}
\normalsize
{\noindent {\bf Notes to Table~\ref{tab:ident}:}}\\
ROXRA1: (7) The IR index ($\alpha_\mathrm{IR}$=dlog($\lambda F_\mathrm{\lambda}$)/dlog$\lambda$) is 
estimated between 2.18--4.69\,$\mu$m from Walter et~al. (\cite{walter94})
and Jensen et~al. (\cite{jensen97}): we find $\alpha_\mathrm{IR}$=-2.5.
(8--9) These values are estimated from Walter $et~al.$ (\cite{walter94}).\\
ROXRA2: (7) Mart{\'\i}n et~al. (\cite{martin98}) classify these stars as CTTS.
(8--9) These values are estimated from GWAYL.\\
ROXRA4: (2) ROXR1-5 was identified with IRS3 and IRS5 (CMFA). \\
ROXRA8: (5) The position of Chini8 (Chini \cite{chini81}) is 33$^{\prime\prime}$ away from this optical star, the good position is given in Table 1 of 
Strom et~al. (\cite{strom95}) by source number 3. (7) The \ISOCAM LW2 and LW3 upper limits exclude an IR excess (see Appendix~D).\\
ROXRA9: (2) ROXR1-12 was identified with an anonymous optical star (CMFA).\\
ROXRA13: (2) ROXR1-15 was neither optical nor IR counterpart (CMFA). \\
ROXRA16: (3) ROXR2-16 was $\sim 1\arcmin$ from ROXRA16 (identified with WSB28; CMFA), ie only two error boxes away, 
but Mart{\'\i}n et~al. (\cite{martin98}) identify ROXR2-16 with another star.\\
ROXRA17: (2) ROXR1-22 was identified with GSS30-IRS1, 2, 3 (CMFA).\\
ROXRA21: (7) S1 is an embedded B-type star (Wilking et~al. \cite{wilking89}; Andr{\'e} et~al. \cite{andre88}).
(8) The visual extinction comes from the data of Lada \& Wilking (\cite{lada84}), see Andr{\'e} et~al. (\cite{andre88}). 
(9) This B3--B5 stellar luminosity is taken from Andr{\'e} et~al. (\cite{andre88}).\\
ROXRA26: (5) In the BKLT survey this well known emission line star appears to be a $6.5\arcsec$ separation binary 
(this can also be suspected from the finding chart): the main component is B162710-241914 ($J$=8.56), 
the second one is B162710-241921 ($J$=11.27).\\
ROXRF1: (5) The IR star B162623-244308 is also in the 90$\%$ confidence error box but its $J$-band luminosity (15.65) is lower than that of the well known emission line star DoAr25=B162623-244311 (9.29).\\
ROXRF7: (2) ROXR1-22 was identified with SR24N and SR24S (CMFA). (5) In our observation \#2 and \#3 the counterpart of ROXRF7 is clearly associated with SR24S. 
In our observation \#1 due to the weakness of the X-ray source the situation is less clear. We associate this source with SR24S.\\
ROXRF9: (2) ROXR1-36 was ambiguously identified with GY193 and GY194 (CMFA).\\
ROXRF12: (5) A weak star is visible in the 90$\%$ confidence error box on the DSS2 image, 
but this star is neither detected by the BKLT survey, nor by the PMM USNO-A1.0 catalogue.\\
ROXRF14: (3) ROXR2-22 was identified with ROXs20A and ROXs20B (CMFA). \\
ROXRF19: (6) This source is just at the border of the \ISOCAM survey. As a part of its flux is lost, Bontemps et~al. (\cite{bontemps00}) 
do not characterize this source. (7) Mart{\'\i}n et~al. (\cite{martin98}) classify these stars as CTTS. We thus consider this source as class II.\\
ROXRF21: (2) ROXR1-43 was identified with GY263 and IRS43 (CMFA). (8) Best value from Grosso et~al. (\cite{grosso97}). 
(9) This is the bolometric luminosity (see Grosso et~al. \cite{grosso97}).\\
ROXRF23: (2) ROXR1-46 was identified with an unnamed optical star (CMFA), probably the star B162730-244726. 
(5) We identified this X-ray with the IR star B162728-244803 just at the border of the \xr error box.\\
ROXRF24: (5) This source is red in the PMM USNO-A1.0 Catalogue (Monet et al. \cite{monet96}) with $B=20.1$ and $R=16.3$.\\
ROXRF25: (2) ROXR1-48 was identified with GY280, GY290, and GY291 (CMFA).\\
ROXRF28: (5) The IR star B162735-244532:B is also in the 90$\%$ confidence error box but its $J$-band luminosity ($>17$) is lower 
than GY296=B162735-244532:A (12.62).\\
ROXRF37: (7) The {\it Hipparcos} distance is 75\,pc: this star is a foreground F2V star.
\vspace{0.2cm}\\
{\noindent {\bf References to Table~\ref{tab:ident}:}
B = Barsony et~al. (\cite{barsony97}).
BBRCG = Barsony et~al. (\cite{barsony89}). 
DoAr = Dolidze \& Arakelyan (\cite{dolidze79}). 
El = Elias (\cite{elias78}). 
GSS = Grasdalen et~al. (\cite{grasdalen73}). 
GY = Greene \& Young (\cite{GY}). 
HD = The Henry Draper catalogue (Draper \cite{draper18}).
IRS = Wilking et~al. (\cite{wilking89}). 
ROXC = Montmerle et~al. (\cite{montmerle83}). 
ROXs = Bouvier \& Appenzeller (\cite{bouvier92}). 
S = abbreviation for ``Source'' in Grasdalen et~al. (\cite{grasdalen73}). 
SKS = Strom et~al. (\cite{strom95}; Table 1). 
SR = Struve \& Rudkj{\o}bing (\cite{struve49}). 
U = The PMM USNO-A1.0 Catalogue (Monet et al. \cite{monet96}). 
VSS = Vrba et~al. (\cite{vrba76}). 
VSSG = Vrba et~al. (\cite{vrba75}). 
WL = Wilking \& Lada (\cite{wilking83}). 
WSB = Wilking et~al. (\cite{wilking87}). 
YLW = Young et~al. (\cite{young86}).}
\end{table*}
\addtocounter{section}{1}

\begin{figure}[!b] 
\resizebox{\hsize}{!}{\includegraphics{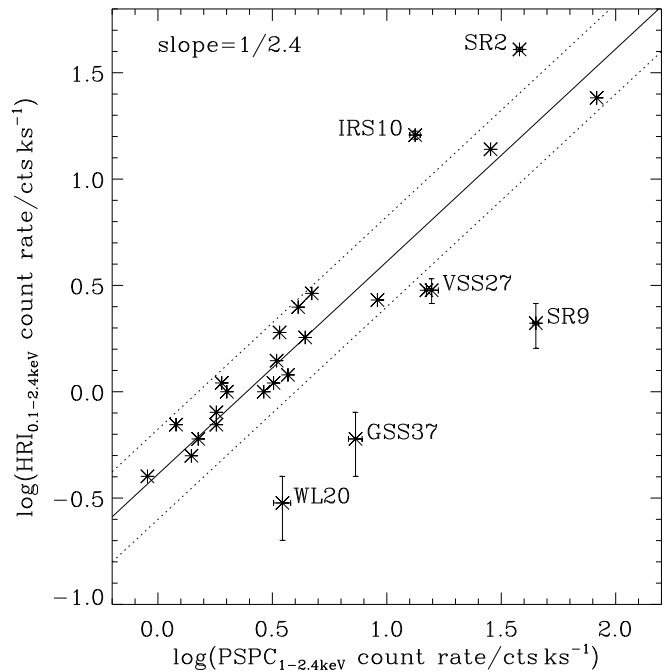}}
\caption{Plot of \ROSAT \HRI (this paper) vs. \PSPC (CMFA) count rates and sources 
variability.
The dashed line is the median value of the conversion factor between \HRI and \PSPC count rates (2.4); 
dotted lines show the dispersion (1\,rms) around this median value.  
Count rate error boxes are shown for sources outward the dotted lines.
Sources below (above) the dotted lines presumably flared during \PSPC (\HRI) observations 
(see text for details).}
\label{HRI-PSPC} 
\end{figure}

Fig.~\ref{HRI-PSPC} displays the \HRI$_{\mathrm{0.1-2.4\,keV}}$
count rate vs. the \PSPC$_{\mathrm{1-2.4\,keV}}$ count rate.  
It shows two classes of sources: 
sources near the median, and sources beyond the median (with error
bars).  The dispersion of points (within 1\,rms) around the median value could be due
to \xr extinction effect on the conversion factor or to a variability factor $\le 2$.  
Preibisch et~al. (\cite{preibisch96})
calculated the conversion for the whole energy band of the \ROSAT \HRI assuming optically
thin plasma emission with $T_{\mathrm X} = 10^7$\,K and different values for the \xr
extinction:  for $N_H$ increasing from $6.5 \times 10^{19}$\,cm$^{-2}$ to
10$^{22}$\,cm$^{-2}$, the conversion factor decreases from 2.5 to 2.0.  Our
observational estimate is in agreement with these values, which also show that
the dependence of the conversion factor on \xr extinction is small compared to
the dispersion of count rates and can be neglected in our plot.  We conclude
that the dispersion is due to variability:  WL20, GSS37, VSS27, and SR9 must have been
in a high state during the \PSPC observation, as were ROXs4 and SR2 during the
\HRI observation, the other sources being essentially unchanged in both
observations.

Using this conversion factor, we can estimate the \HRI count rate from the \PSPC
count rate, and compare it with our \HRI threshold computed with EXSAS.  We find
that 14 sources are below this threshold (ROXR1-1, 6, 7, 16, 20, 21, 27,
33, 34, 37, 40, 53, and ROXR2-16, 18), and the other 7 sources were in a
high state during the \PSPC observation (ROXR1-19, 30, 45, 47, and ROXR2-27,
30, 33).  We conclude that the non-detection of the 21 \PSPC sources by the \HRI
can be fully explained by the difference in sensitivity and intrinsic
variability. 

We can also compare our detections with \ASCA (Koyama et~al. \cite{koyama94}; Kamata et~al.
\cite{kamata97}; see observation field in Fig.~\ref{fields-map}) despite its lower
angular resolution.  Kamata et~al. (\cite{kamata97}) detected 19 \xr sources, of which 10
were previously observed by Koyama et~al. (\cite{koyama94}).  Compared to \Einstein and
\ROSAT \PSPC observations, 7 new \xr sources were discovered by {\it ASCA}.\footnote{Note that 
\xr source 9B of Kamata et~al. (\cite{kamata97}) is in fact ROXR1-45
of CMFA.} These 7 \xr sources are also not detected with the {\it HRI}. 
On the 12 sources already observed by \Einstein and {\it ROSAT PSPC}, 
we detect 9 sources, the 3 others being below our sensitivity threshold according to 
the above conversion factor.

\section{Optical/IR counterpart without IR classification}
\label{appendix:sed}

\begin{figure}[!ht]
\resizebox{\hsize}{!}{\includegraphics{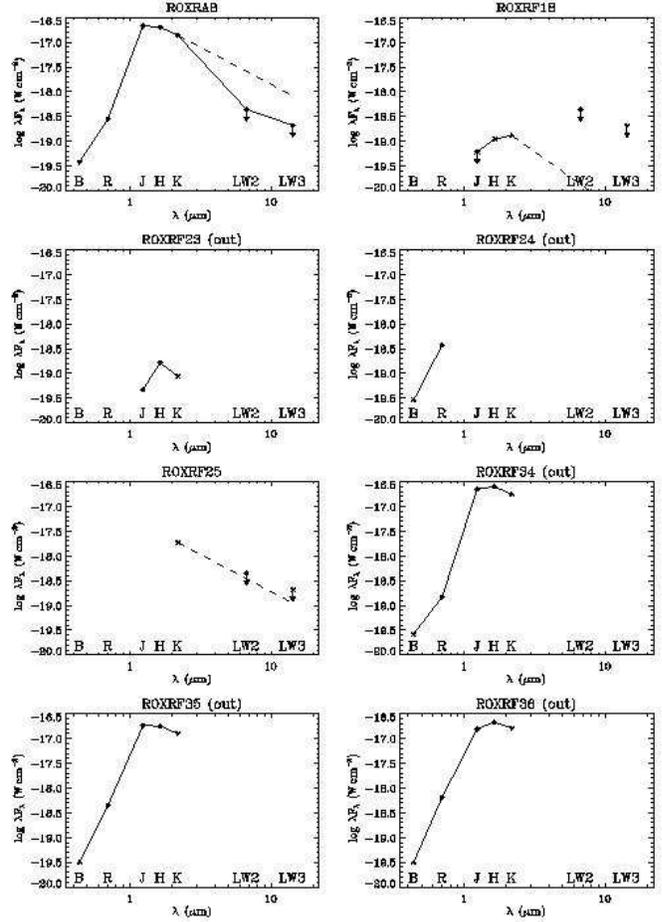}}
\caption{Spectral Energy Distribution of optical/IR counterparts without IR classification. 
B and R data come from the PMM USNO-A1.0 catalogue (Monet et al. \cite{monet96}); $J$, $H$, $K$ data come 
from BKLT; LW2 and LW3 are the \ISOCAM upper limit at 10\,mJy from Bontemps et~al. (\cite{bontemps00}). 
The dashed line shows the limit between Class~II source and Class~III source classification 
($\alpha_\mathrm{IR}$=dlog($\lambda F_\mathrm{\lambda}$)/dlog$\lambda$=-1.5 from AM).
``out'' means that the source is outside the \ISOCAM survey.}
\label{sed}
\end{figure}

Nine X-ray sources have optical/IR counterpart for which the IR classification is not known.
Three of these X-ray sources are found in the \ISOCAM survey, but with only upper limits 
in LW2 and LW3 filters.
We give here their spectral energy distribution (see Fig.~\ref{sed}), 
and discuss their possible IR classification.
In case of doubt, the resulting Class~II (or Class~III) source candidates have not been included 
in the statistic studies of this article.

{\bf ROXRA8:}
The counterpart of this \xr source is the optical star Chini8=SKS3 ($R=16.6$, $K=9.5$).
The \ISOCAM LW2 and LW3 upper limits exclude an IR excess. 
We classify this source as a Class~III source.

{\bf ROXRF12:}
A weak star is visible in the 90$\%$ confidence error box on the DSS2 image, 
but this star is neither found in the BKLT survey, nor in the PMM USNO-A1.0 catalogue.
The low optical/near IR magnitudes imply a low luminosity for this object.
We propose this source as a weak Class~II or Class~III source candidate detected during 
a strong \xr flare. 
This source may also be a brown dwarf.

{\bf ROXRF18:}
The counterpart of this \xr source is the IR star B162720-243820 ($K=14.6$).
This star is visible in the DSS2 (red) optical image, but it is not in the 
PMM USNO-A1.0 catalogue (probably because only stars appearing both in blue and
red images were accepted), thus we have no estimate of its B and R magnitudes.
The low near IR magnitudes imply a low luminosity for this object.
This source may be a weak Class~II or Class~III source candidate detected during a strong \xr flare.
\ISOCAM LW2 and LW3 upper limits do not exclude an IR excess for this object.
This object can also be a weak Class~I protostar with a strong \xr flare. 

{\bf ROXRF23:}
The counterpart of this \xr source is the IR star B162728-244803
$(K=14.1$; see note in Table~\ref{tab:ident}). 
The low near IR magnitudes imply a low luminosity for this object.
The SED of this source peaks in the $H$-band.
We propose this source as a weak Class~II or Class~III source candidate detected during 
a strong \xr flare.

{\bf ROXRF24:}
The counterpart of this \xr source is an optical star ($R=16.3$) named only in 
the PMM USNO-A1.0 catalogue (Monet et al. \cite{monet96}). 
Unfortunately, this object lies outside the BKLT survey.
We propose this source as Class~II or Class~III source candidate.

{\bf ROXRF25:}
The counterpart of this source is BBRCG50 observed only in $K$-band ($K=11.7$). 
This source is not retrieved in BKLT. 
The \ISOCAM LW2 and LW3 upper limits exclude a strong IR excess.
We propose this source as Class~II or Class~III source candidate.

{\bf ROXRF34:}
The counterpart of this \xr source is the optical star WSB58=B162800-244819 ($R=17.3$, $K=9.3$).
Wilking et~al. (\cite{wilking87}) noted a probable H$_\alpha$ detection needing confirmation. 
The SED of this source peaks in the $H$-band.
We propose this source as Class~II or Class~III source candidate.

{\bf ROXRF35:}
The counterpart of this \xr source is the optical star B162800-245340 ($R=16.1$, $K=9.6$). 
The SED of this source peaks in the $J$-band.
We propose this source as Class~II or Class~III source candidate.

{\bf ROXRF36:}
The counterpart of this \xr source is the optical star B162812-245043 ($R=15.7$, $K=9.4$), 
which appears to be a close binary ($\sim 1.5\arcsec$) in the second Digitized Sky Survey 
(see Fig.~\ref{finding_map}). 
The SED of this source peaks in the $H$-band.
We propose this source as Class~II or Class~III source candidate.


\end{document}